\documentclass[11pt]{article}
\usepackage{graphicx}
\usepackage{amssymb}
\usepackage{amscd}
\usepackage{mathrsfs}
\usepackage{longtable,lscape}
\usepackage{amsthm}
\usepackage{amsfonts}
\usepackage{amsmath}
\usepackage{bbm}
\usepackage{float}
\usepackage{url}
\usepackage[title]{appendix}
\usepackage{csquotes}

\newcommand{\captionfonts}{\footnotesize}
\makeatletter  
\long\def\@makecaption#1#2{%
  \vskip\abovecaptionskip
  \sbox\@tempboxa{{\captionfonts #1: #2}}%
  \ifdim \wd\@tempboxa >\hsize
    {\captionfonts #1: #2\par}
  \else
    \hbox to\hsize{\hfil\box\@tempboxa\hfil}%
  \fi
  \vskip\belowcaptionskip}
\makeatother 
\begin{document}
\title{The cognitive triple-slit experiment}
\author{Luca Sassoli de Bianchi$^1$, Massimiliano Sassoli de Bianchi$^{1,2}$\vspace{0.5 cm} \\ 
        $^1$ Laboratorio di Autoricerca di Base \\ 
        \normalsize\itshape
         Via Cadepiano 18, 6917 Barbengo, Switzerland \\
        \normalsize
        E-Mail: \url{sdb.luca@protonmail.ch} 
        \vspace{0.5 cm} \\ 
        \normalsize\itshape
             \normalsize\itshape
        $^2$ Center Leo Apostel for Interdisciplinary Studies, 
         Brussels Free University \\ 
        \normalsize\itshape
         Krijgskundestraat 33, 1160 Brussels, Belgium \\
        \normalsize
        E-Mail: \url{msassoli@vub.ac.be}
          \vspace{0.5 cm} \\ 
         \normalsize\itshape
              }
\date{}
\maketitle
\begin{abstract}
\noindent 
Quantum cognition has made it possible to model human cognitive processes very effectively, revealing numerous parallels between the properties of conceptual entities tested by the human mind and those of microscopic entities tested by measurement apparatuses. The success of quantum cognition has also made it possible to formulate an interpretation of quantum mechanics, called the conceptuality interpretation, which ascribes to quantum entities a conceptual nature similar to that of human concepts. The present work fits into these lines of research by analyzing a cognitive version of single-slit, double-slit, and triple-slit experiments. The data clearly show the formation of the typical interference fringes between the slits, as well as the embryos of secondary fringes. Our analysis also shows that while quantum entities and human concepts may share a same conceptual nature, the way they manifest it in specific contexts can be quite different. This is also evident from the significant deviation from zero observed for the Sorkin parameter, indicating the presence of strong irreducible third-order interference contributions in human decision.
\end{abstract}

\section{Introduction\label{intro}}

Young's double-slit experiment, first performed in the early 1800s, is probably the most famous experiment in physics \cite{thomas1804, taylor1909}, which has been repeated over the decades not only with photons but also with electrons \cite{jonsson1961,merli1974,rosa2012}, neutrons \cite{zeilinger1988}, atoms \cite{carnal1991} and even large molecules \cite{fein2019}. The experiment has played an important role in the debates on the interpretation of quantum mechanics, and even Albert Einstein and Niels Bohr often focused on it in the course of their historical discussions on the completeness of quantum mechanics \cite{jammer1974}. A particularly accurate and didactic analysis of it can be found in the famous Feynman Lectures on Physics \cite{feynman1964}. The experiment notably revealed that the interference effects resulting from the quantum superposition principle are not the expression of a collective effect, considering that the interference fringes that form on the detection screen appear even when a single quantum entity at a time interacts with the apparatus \cite{feynman1964}. This means that while the localized impacts on the detection screen seem to indicate a particle-like nature, the creation of the interference pattern also reveals that what crosses the double-slit screen is more like a wave phenomenon. However, since a physical entity cannot be both a wave and a particle, these experiments emphasize that the nature of a microscopic entity is of an even different nature.

\begin{figure}[htbp]
\begin{center}
\includegraphics[width=13cm]{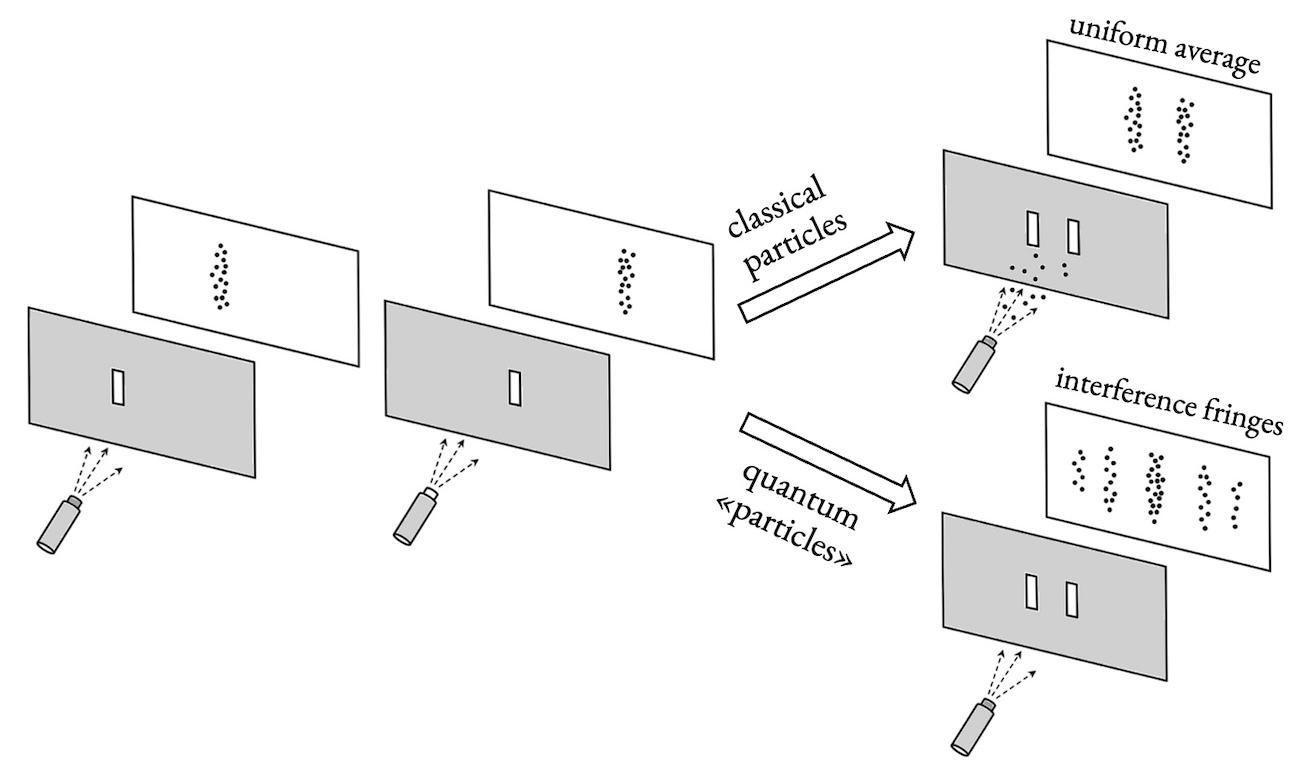}
\caption{A schematic representation of the two-slit experiment in the case of classical (top right) and quantum (bottom right) entities.}
\label{classicalquantum}
\end{center}
\end{figure}

For classical particles, the distribution of impacts when both slits are open can be reconstructed as a uniform average of the distributions with only one slit open at a time. In the quantum case, however, this does not hold, due to interference effects. This becomes particularly clear when one observes that the main fringe is located in the center of the two slits, where the probability of an impact is classically the lowest; see Fig.~\ref{classicalquantum}. The formation of an interference fringe corresponds to a region on the screen where the probability is \emph{overextended} with respect to the classical average (constructive interference), while the regions where no impacts are found are those where the probability is \emph{underextended} (destructive interference).

Cognitive analogues of the double-slit experiment have also been proposed \cite{aerts2009}. We will examine a significant example in detail, as this will clarify the design of the cognitive experiment presented in Section~\ref{doubleslitcognitive}. In the 1980s, psychologist James Hampton conducted an experiment in which 40 students were presented with $24$ exemplars of \emph{Food} and asked to what extent they thought they were related (in terms of membership and typicality) to the concepts: (a) \emph{Fruit}, (b) \emph{Vegetable}, and (c) \emph{Fruit or vegetable} \cite{Hampton1988}. The  $24$ exemplars of \emph{Food} played the same role as the different locations where the trace of an impact can be observed on the detection screen of a double-slit experiment, with the two abstract concepts \emph{Fruit} and \emph{Vegetable} playing the role of the two slits. More precisely, choosing a typical exemplar of \emph{Fruit} is here equivalent to choosing a good example of an impact of an entity passing through the first slit. Similarly, selecting a typical exemplar of \emph{Vegetable} is equivalent to providing a good example of the impact of an entity passing through the second slit. Finally, selecting a typical exemplar of  \emph{Fruit or vegetable} is equivalent to providing a good example of an impact of an entity passing through both slits, i.e., through the first slit \emph{or} the second slit. 

\begin{figure}[htbp]
\begin{center}
\includegraphics[width=13cm]{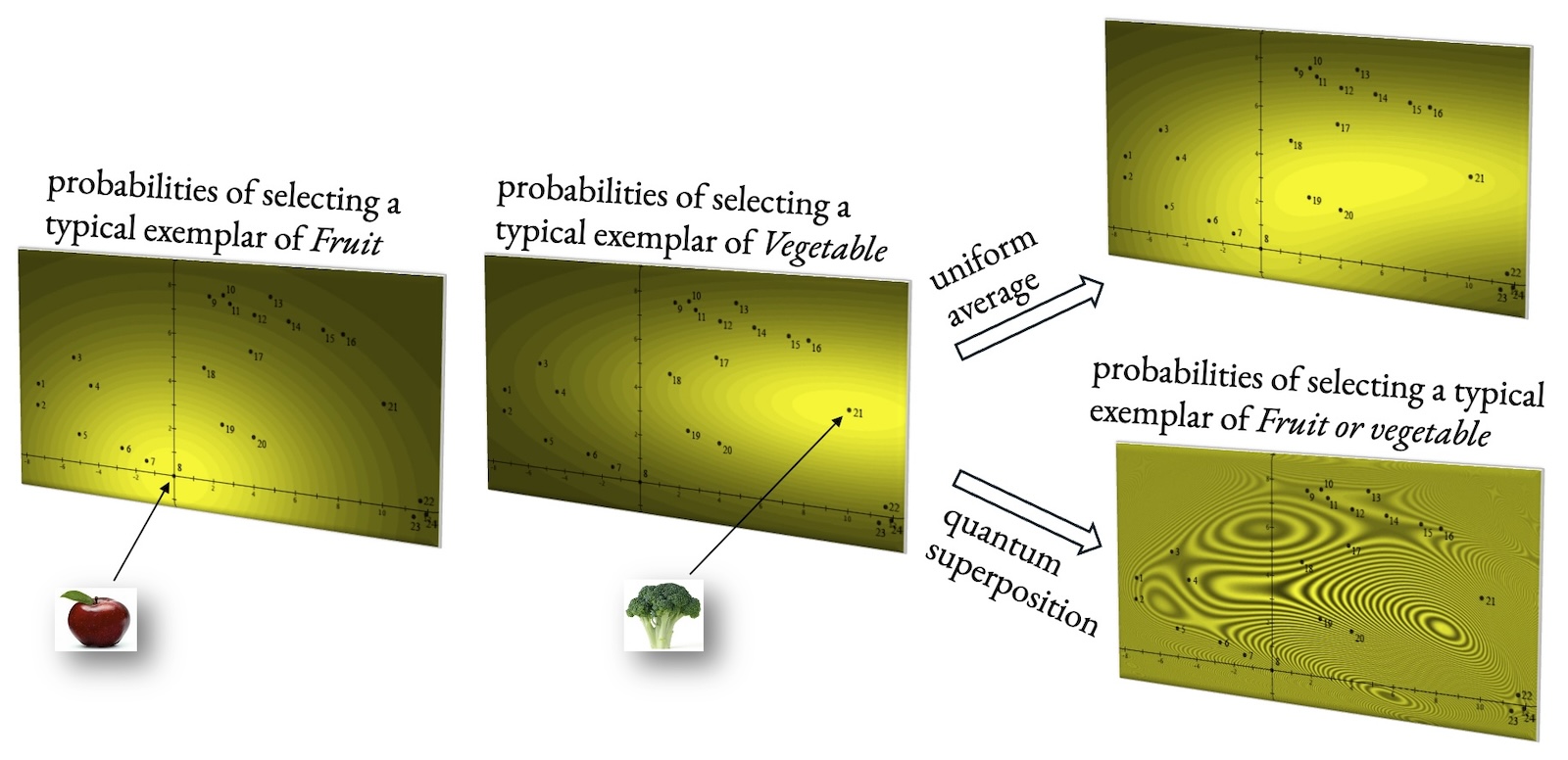}
\caption{(color online) On the left are the experimental probabilities of selecting a typical exemplar of  \emph{Fruit} and of  \emph{Vegetable}, with \emph{Apple} and \emph{Broccoli} being the most frequently chosen, respectively. The top right figure shows their uniform average, which differs greatly from the experimental probabilities of selecting a typical exemplar of  \emph{Fruit or vegetable}, modeled using the quantum superposition principle (bottom right figure) and exhibiting birefringence-like interference phenomena \cite{aerts2009}.}
\label{birefringence}
\end{center}
\end{figure}

An expectation based on classical prejudice would tell us that when faced with question (c), about the combination \emph{Fruit or vegetable}, the probability of selecting a given exemplar of \emph{Food} would be the uniform average of the probabilities describing the situations of questions (a) and (b), but this is not what Hampton's data revealed. Similar to the constructive and destructive interference effects in the impact statistics of typical double-slit experiments, overextensions and underextensions of the probabilities were observed, and when Hampton's data were represented in a quantum-like way, a complex interference-like pattern was revealed, reminiscent of those obtained in the phenomena of birefringence \cite{aerts2009}; see Figure~\ref{birefringence}.

More precisely, to obtain this pattern, Aerts represented the concepts \emph{Fruits}, \emph{Vegetables} and \emph{Fruits or vegetables} by unit vectors $|F\rangle$, $|V\rangle$ and $|FV\rangle ={1\over\sqrt{2}}(|F\rangle +|V\rangle)$, respectively. He then chose them so that the wave functions $\langle x|F\rangle$ and $\langle x|V\rangle$ would be two-dimensional, with a Gaussian real part. Placing all the exemplars on a two-dimensional plane, in such a way that, for their coordinates, the quantum probabilities $P_F(x)=|\langle x|F\rangle|^2$, $P_V(x)=|\langle x|V\rangle|^2$ and $P_{FV}(x)=|\langle x|FV\rangle|^2$ properly modeled the experimental probabilities, he then obtained the pattern shown in Figure~\ref{birefringence}  \cite{aerts2009}.  

\begin{figure}[htbp]
\begin{center}
\includegraphics[width=8cm]{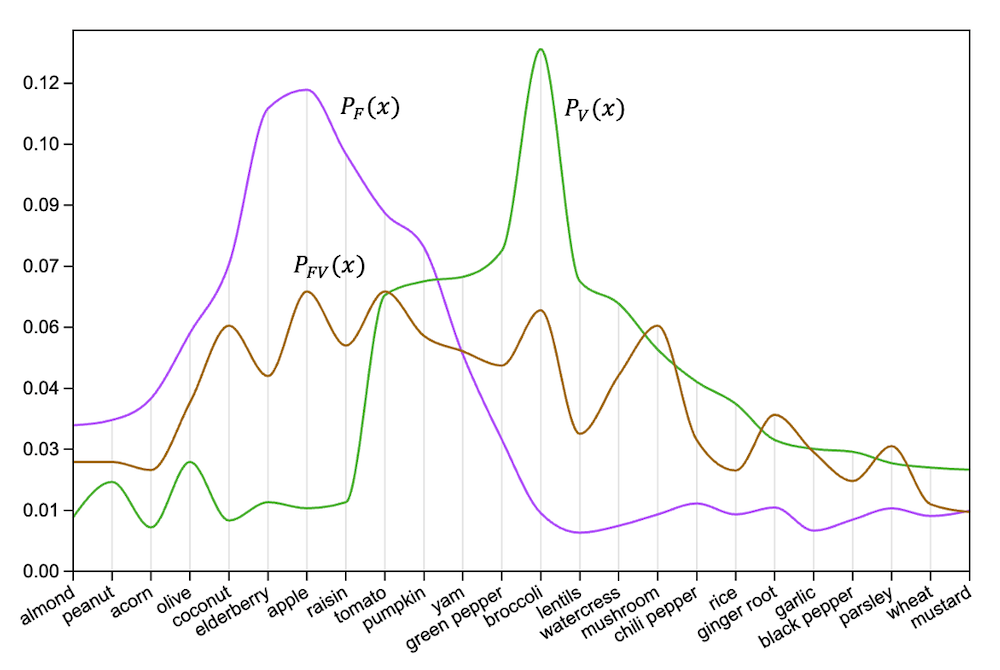}
\caption{(color online) A continuous approximation of the discrete single-slit functions $P_F(x)$ (purple color) and $P_V(x)$ (green color), and corresponding double-slit function $P_{FV}(x)$ (brown color), which is the one-dimensional equivalent of the two-dimensional pattern exhibited in Figure~\ref{birefringence}.}
\label{Figure-FV-onedimensional}
\end{center}
\end{figure}

Note that the use of Gaussians requires at least two dimensions to distribute the different datapoints. This does not mean that it is not possible to also describe Hampton's data using a one-dimensional screen of exemplars. For this, one has to position them in such a way that two clear peaks appear for the two single-slit experiments relating to the concepts \emph{Fruits} and \emph{Vegetables}, centered on the exemplars that obtained the highest score. Obviously, this requires giving up perfectly symmetrical Gaussian distributions. Once this is done, one can deduce the two-slit curve for the combination \emph{Fruits or Vegetables}, as shown in Figure~\ref{Figure-FV-onedimensional}. 

A possible interpretation of the quantum-like\footnote{As Figure~\ref{Figure-FV-onedimensional} shows, the inequality $P_{FV}(x) \ge \max\{P_F(x), P_V(x)\}$ is violated by many exemplars. Therefore, Hampton's disjunction effects cannot be consistently modeled using a single classical probability space. The same will be true for the probabilities obtained in our experiment. See for example the analysis in \cite{aerts2009b}.} overextension and underextension effects obtained in Hampton's experiment is that they would correspond to regions minimizing (respectively, maximizing) the doubt of the participants as to how to classify a given exemplar. Let us explain this in some detail. When the students were asked about a typical exemplar of \emph{Fruit}, the most frequent answer (the one receiving the higher rating) was \emph{Apple}, and when asked about a typical exemplar of \emph{Vegetable}, it was \emph{Broccoli}. On the other hand, \emph{Mushroom}, although considered to be more representative of \emph{Vegetable} than of  \emph{Fruit}, it wasn't considered a typical exemplar of either, in the sense that the associated probabilities were very small compared to those associated with \emph{Apple} and \emph{Broccoli}. But when the students were asked about a typical exemplar of the conceptual combination \emph{Fruit or vegetable}, it was \emph{Mushroom} that received the higher score, with a probability that was almost double compared to the uniform average of the probabilities related to only  \emph{Fruit} and only \emph{Vegetable} (overextension effect). This deviation can be explained by observing that the disjunction \emph{Fruit or vegetable} is a new emerging concept, not reducible, in regard to its meaning, to the meanings of its components taken individually. More precisely, \emph{Fruit or vegetable}, in addition to its pure logical meaning, also conveys the meaning of a doubt as to whether a given exemplar of \emph{Food} can be classified as  \emph{Fruit} or as  \emph{Vegetable}, and since \emph{Mushroom} does not classify well neither as \emph{Fruit} nor as \emph{Vegetable}, it will be an excellent exemplar to be classified as \emph{Fruit or vegetable}, hence the overextension effect. Similarly, an exemplar like  \emph{Elderberry}, since it is typical of  \emph{Fruit} and not typical of \emph{Vegetable}, and is scored accordingly when answering questions (a) and (b), it is not a good example of a situation of doubt as to whether it belongs to one of these two categories. So, it will receive a score lower than the uniform average of the probabilities related to only  \emph{Fruit} and only \emph{Vegetable}, hence the underextension effect. 

If we follow this line of reasoning, we can try to provide a cognitive-like interpretation of the double-slit experiment by claiming that the fringes of higher (lower) intensity in the interference pattern correspond to regions that maximize (minimize) doubt about the slits from which the impact came from. In other words, an impact-rich fringe would be a region on the screen that makes it difficult to guess which slit the quantum entity passed through, while it would be exactly the opposite for the regions on the screen with a low density of impacts \cite{aerts2009}. 

This analogy with Hampton's experiment requires interpreting \emph{Fruit} and \emph{Vegetable} as two possible states of the concept \emph{Food}. The conceptual combination \emph{The entity passes through slit-1} is then the analogue of the conceptual combination \emph{The food is a fruit}, the conceptual combination \emph{The entity passes through slit-2} is the analogue of \emph{The food is a vegetable}, and the conceptual combination \emph{The entity passes through slit-1 or slit-2} is the analogue of  \emph{The food is a fruit or vegetable}. The selected exemplars also correspond to possible states of the \emph{Food} conceptual entity, such as the state \emph{The food is an apple}. These are the analogues of the states corresponding to the different locations where the quantum entity can be detected on the screen, such as \emph{The entity is located at position $x=0$}. Hence, instead of saying ``the entity passing through slit-1 (respectively, slit-2) is detected in position $x$,'' we should say ``position $x$ is a good example of the entity passing through slit-1 (respectively, slit-2).'' Similarly, instead of saying ``the entity passing through slit-1 or slit-2 is detected in position $x$,’’ we should say ``position $x$ is a good example of the entity passing through slit-1 or slit-2.''

Following the above analogy, which is rooted in the \emph{conceptuality interpretation of quantum mechanics} (first proposed by Diederik Aerts in 2009 \cite{aerts2009} and further developed by his Brussels group \cite{aertssassoli2018, aertssassolisozzoveloz2019, aertsetal2020, sassoli2021, aertssassoli2022, aertssassoli2024, aertssassolisozzoveloz2024a, aertssassolisozzoveloz2024b}), the detection screen must be viewed as a cognitive structure sensitive to the meaning carried by the entities interacting with it. Outcomes are thus selected in a manner analogous to how cognitive entities respond to questions. In other words, within this interpretation, quantum entities are assumed to be conceptual rather than spatiotemporal, and this shift in perspective explains their apparently paradoxical behavior (human concepts and the conceptual entities of the microworld should however not be conflated, just as electromagnetic and acoustic waves should not be confused, despite both sharing an undulatory nature). 

In the double-slit experiment, when the state expresses a lack of knowledge, i.e., incomplete information about the slit through which the entity passes, this incompleteness is associated, in the conceptual domain where meaning-driven interactions occur, with a genuine new element of reality. Hence, in the quantum formalism, the situation is described by a superposition state rather than a mixture. The central fringe then becomes the region where the best answers to the question posed are located, in the sense of providing the most faithful exemplifications of an ontic, not merely epistemic, condition of indeterminacy.

It is worth mentioning that in the last two decades a field of research called \emph{quantum cognition} has emerged, using quantum mechanics, and especially its mathematics, to find new tools for explaining and modeling human cognition, especially how meaning is assigned to concepts and their combinations, and how decision-making processes occur depending on context. It is because of the success of this quantum approach to human modeling that the speculative hypothesis at the basis of the conceptuality interpretation subsequently emerged, according to which this success was not accidental, but due to the fact that \emph{quantumness} and \emph{conceptuality} are notions that point to a same nature. Indeed, by using the quantum formalism and the notions that it brings to bear (in particular, \emph{potentiality} and \emph{context-dependence}), quantum cognition was able to bring to the fore, in the psychological laboratories, all those effects that are typical of quantum entities, such as superposition and interference effects, entanglement, complementarity and indistinguishability \cite{aertssassolisozzoveloz2024a, aertssassolisozzoveloz2024b,Khrennikov2010, BusemeyerBruza2012,HavenKhrennikov2013,Wendt2015,aertsetal2019,aertsbeltran2020,aertsaerts2022}. 

Positioning itself at the frontier between the conceptuality interpretation and the research in quantum cognition, this article explores to what extent the human mind is able to reproduce a fringe-like pattern (or its embryo) reminiscent of those observed in physics laboratories, when confronted to an experiment that mimics the double and triple-slit experimental situations as closely as possible. More precisely, the article is organized as follows. In Section~\ref{doubleslit}, we recall how the quantum formalism describes the two and three-slit situations. For the latter, we derive the so-called {\it Sorkin parameter}, showing that its value is zero when the state of the system is a clear-cut superposition of the different possibilities at play, meaning that there are no genuine third-order interference contributions. In Section~\ref{third}, we explain why, contrary to what has been initially believed, the quantum formalism also predicts the existence of genuine third-order interferences, albeit in typical experimental settings they are of a much lower magnitude than the second-order ones. In Section~\ref{doubleslitcognitive}, we describe the cognitive single-slit, double-slit and triple-slit experiments that we have carried out, and in Section~\ref{analysisdata} we provide a detailed analysis of the collected data. Finally, in Section~\ref{conclusion}, we offer some concluding reflections.

\section{The quantum triple-slit experiment\label{doubleslit}}

For classical, corpuscular entities, the average of the two single-slit situations reproduce the double-slit situation. More precisely, denoting $P_{\rm cl}(x|12)$ the (classical) probability of observing an impact at point $x$ of the detection screen, when both slits are open, we can always write it as the uniform average $P_{\rm cl}(x|12)= \tfrac{1}{2}P_{\rm cl}(x|1) + \tfrac{1}{2}P_{\rm cl}(x|2)$, where we have assumed that the probabilities of passing through slit-$1$ and slit-$2$ are the same, and $P_{\rm cl}(x|1)$ and $P_{\rm cl}(x|2)$ are the probabilities of observing an impact at point $x$ when only slit-$1$ and slit-$2$ are open, respectively. As mentioned in the previous section, the above average does not hold anymore if the entities are quantum. More precisely, if $\psi(x|1)$ and $\psi(x|2)$ are the wave functions at the detection screen, assumed here to be orthogonal unit vectors,\footnote{The orthogonality of the one-slit wave functions can be reasonably assumed because the slits correspond to spatially disjoint regions. Therefore, the wave functions can be considered orthogonal at the slit plate level, since their supports do not overlap. Since unitary evolution preserves orthogonality, the wave functions will also remain orthogonal at the detection screen level, even though they overlap at the screen and produce interference contributions. (Orthogonality in Hilbert space does not forbid overlap in configuration space.)} describing the situations where the quantum entity passed through slit-$1$ and slit-$2$ only, respectively, then we know, according to the Born rule, that the associated quantum probabilities are given by the squared modules $P(x|1)= |\psi(x|1)|^2$ and $P(x|2)= |\psi(x|2)|^2$. When both slits are open, we can apply the prescription saying that when there are alternatives (slit-$1$ or slit-$2$), the probability amplitude, $\psi(x|12)$, is obtained from the normalized sum of the probability amplitudes associated with the alternatives considered separately: $\psi(x|12)=\tfrac{1}{\sqrt{2}}\psi(x|1)+\tfrac{1}{\sqrt{2}}\psi(x|2)$. Therefore, the associated probability is:
\begin{eqnarray}
P(x|12)&=&|\psi(x|12)|^2 = |\tfrac{1}{\sqrt{2}} \psi(x|1)+\tfrac{1}{\sqrt{2}}\psi(x|2)|^2 \nonumber\\
&=& \tfrac{1}{2}P(x|1) + \tfrac{1}{2}P(x|2) + \Re\, [\psi(x|1)^*\psi(x|2)]=\overline{P}(x|12)+I(x|12)
\label{superposition}
\end{eqnarray}
where we have defined the uniform average $\overline{P}(x|12)\equiv \tfrac{1}{2}P(x|1) +\tfrac{1}{2}P(x|2)$, and the two-slit interference contribution $I(x|12) \equiv  \Re\, [\psi(x|1)^*\psi(x|2)]$, which accounts for the overextension and underextension effects. 

In the situation with three slits one has to consider the superposition: 
\begin{equation}
\psi(x|123)=\tfrac{1}{\sqrt{3}} \psi(x|1)+\tfrac{1}{\sqrt{3}}\psi(x|2)+\tfrac{1}{\sqrt{3}}\psi(x|3)
\label{superposition2}
\end{equation}
and the associated probability is:
\begin{eqnarray}
P(x|123)&=& |\psi(x|123)|^2 = \tfrac{1}{3}P(x|1) + \tfrac{1}{3}P(x|2) + \tfrac{1}{3}P(x|3) \nonumber\\
&+& \tfrac{2}{3}  \Re\, [\psi(x|1)^*\psi(x|2)]+ \tfrac{2}{3} \Re\, [\psi(x|2)^*\psi(x|3)]+\tfrac{2}{3}\Re\, [\psi(x|1)^*\psi(x|3)]\nonumber\\
&=& \overline{P}(x|123)+\tfrac{2}{3}\left[I(x|12)+I(x|23)+I(x|13)\right]
\label{prob123-quantum2}
\end{eqnarray}
where we have defined the uniform average  $\overline{P}(x|123) \equiv \tfrac{1}{3}\sum_{i=1}^3P(x|i)$, and the two-slit interference contributions $I(x|12) \equiv  \Re\, [\psi(x|1)^*\psi(x|2)]$, $I(x|23) \equiv   \Re\, [\psi(x|2)^*\psi(x|3)]$ and $I(x|13) \equiv  \Re\, [\psi(x|1)^*\psi(x|3)]$. By definition of the latter, we can also write $I(x|12)=P(x|12)-\overline{P}(x|12)$, $I(x|23)=P(x|23)-\overline{P}(x|23)$, and $I(x|13)=P(x|13)-\overline{P}(x|13)$. Inserting these three expressions in (\ref{prob123-quantum2}), then observing that, by definition of uniform averages, we also have 
\begin{equation}
\overline{P}(x|123)-\tfrac{2}{3}\left[\overline{P}(x|12)+\overline{P}(x|23)+\overline{P}(x|13)\right]=-\tfrac{1}{3}\overline{P}(x|123)
\end{equation}
one obtains the identity: 
\begin{equation}
P(x|123)=\tfrac{2}{3}[P(x|12)+P(x|23)+P(x|13)]-\tfrac{1}{3}[P(x|1)+P(x|2)+P(x|3)]
\label{prob123-quantum2bis}
\end{equation}
In other words, defining the quantity:
\begin{equation}
\epsilon(x|123)\equiv P(x|123)-\tfrac{2}{3}[P(x|12)+P(x|23)+P(x|13)]+\tfrac{1}{3}[P(x|1)+P(x|2)+P(x|3)]
\label{sorkin-uniform}
\end{equation}
called the \emph{Sorkin parameter}  \cite{sorkin1994}, one obtains the condition: $\epsilon(x|123)=0$, for all $x$. So, the Sorkin parameter, an expression consisting only of experimentally accessible probabilities, quantifies possible deviations of the three-slit interference contribution from contributions only coming from second-order two-slit interference effects. Indeed, inserting $I(x|123)\equiv P(x|123)-\overline{P}(x|123)$ in (\ref{prob123-quantum2}), it can also be written
\begin{equation}
\epsilon(x|123)=I(x|123)-\tfrac{2}{3}[I(x|12)+I(x|23)+I(x|13)]
\label{sorkin-uniform2}
\end{equation}
Quantum mechanics predicts that the Sorkin parameter is zero, provided one can interpret the probabilities $P(x|i)$, $i=1,2,3$, as describing the single-slit experiments, i.e., that the prescription (\ref{superposition2}) is correct, which however is not generally true, as we will discuss in the next section. Note that $\epsilon(x|123)$ is trivially zero in classical mechanics, as all interference contributions are then zero.

\section{Third-order interference effects\label{third}}

As first pointed out by Rafael Sorkin \cite{sorkin1994}, the Born rule being quadratic in the wave function it does not allow for genuine third-order interference contributions \cite{ududec2011,nyman2011}. From this follows that the Sorkin parameter (\ref{sorkin-uniform}) is zero, which is something that can be experimentally tested. The first experiments to measure the Sorkin parameter were performed in 2009 by Urbasi Sinha and colleagues \cite{sinhaetal2009,sinhaetal2010,sinha2011}. Their result was that, within their error margins, the nullity of the Sorkin parameter had been verified, and in this sense the prediction of the Born rule had been confirmed. More precisely, the magnitude of a third-order interference contribution was shown to be less than $10^{-2}$ of the magnitude of the expected second-order interference contributions. 

Note that any setting that allows for the presence of three mutually exclusive paths can be used to test the nullity of the Sorkin parameter. This was the situation of the experiment performed by S\"{o}llner and colleagues \cite{sollneretal2012} with a three-path photonic interferometer, which allowed them to improve by about an order of magnitude the best previous experiments. 

\begin{figure}[htbp]
\begin{center}
\includegraphics[width=8cm]{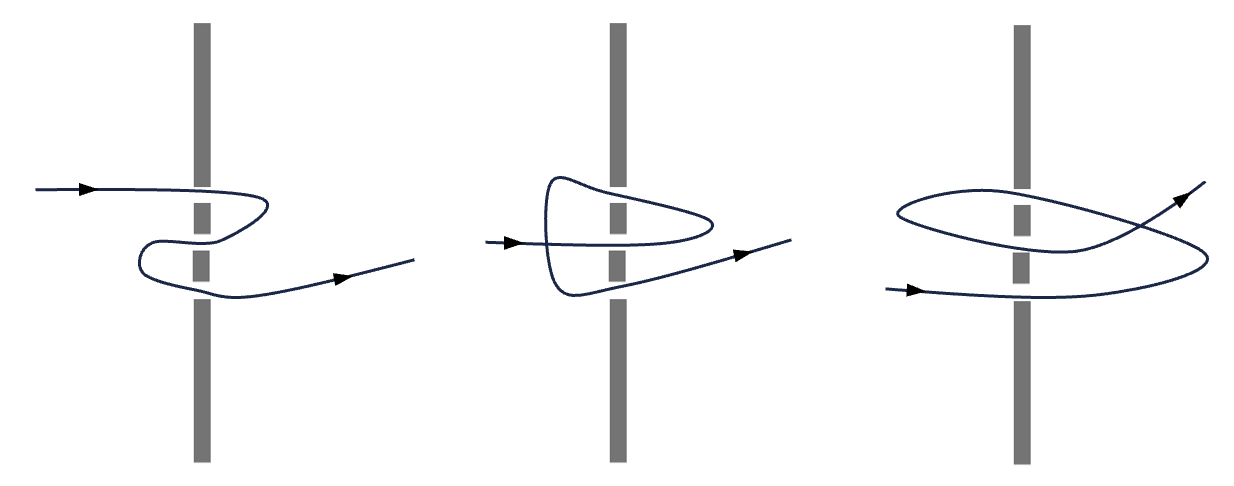}
\caption{Examples of exotic paths that cross all three slits. See the analysis in \cite{yabuki1986}.}
\label{exoticpaths}
\end{center}
\end{figure}

Up to this point, it was believed that Sorkin's initial analysis and subsequent experiments were to be interpreted as confirming the peculiarity of the Born rule in not predicting genuine third-order interference contributions. But things were not as they appeared. As pointed out in \cite{deraedt2012}, it was incorrect to assume that it is always possible to write the wave function on the detection screen, when all three slits are open, as the superposition (\ref{superposition2}) of contributions due only to the three alternatives in play, i.e., to passing through only one of the three slits. One way to understand this is to observe that, due to the phenomenon of the spreading of the wave-packet, it is not possible to decompose a wave-function on the screen as a superposition of wave-functions uniquely associated with the single-slit situations. In fact, during its evolution in time, the support of any wave function will inevitably always extend to touch with its tails all the slits. This can be expressed in a more specific way by considering a Feynman path-integral approach \cite{feynman1965}, which involves integration over all possible paths, including those passing through several slits (forming loops for example; see Fig.~\ref{exoticpaths}) before reaching the detection screen \cite{yabuki1986}, and this is another way of saying that the wave function is not uniquely decomposable as a superposition of contributions attributable to individual slits. While these looped paths contribute less to the overall probability than paths closer to classical trajectories, they cannot be neglected. In other words, contrary to the initial belief, a non-zero value of the Sorkin parameter was not to be considered as a violation of the Born rule \cite{deraedt2012, sinhatal2015, quach2017,sinhaetal2014}. To explain it yet another way, Yabuki emphasized that when we close slits we modify the Hamiltonian. Thus, in a double-slit problem the solutions with only one slit open are not directly related to the solutions with both slits open, since they correspond to different physical contexts, so the superposition principle cannot be applied naively \cite{yabuki1986}. And of course the same is true for a three-slit experiment. 

As we said, a third-order interference contribution arising from exotic paths through the three slits will typically be very small and difficult to detect experimentally as a non-zero Sorkin parameter. Nevertheless, in 2016, Maga\~{n}a-Loaiza and colleagues were able to show for the first time such a contribution to the formation of optical interference fringes \cite{maga2016}. Note, however, that there are experiments, such as the extended Mach-Zehnder interferometer experiment described in \cite{franson2010}, for which these contributions are not expected, because the experiments are designed to probe precisely the interference of three waves, propagating along non-overlapping pathways. And the same is true for regimes where the operating wavelengths are much smaller than the distance between the slits. In these situations, one can still write the general solution of the problem with three open slits as an unambiguous superposition of three amplitudes, each of which can be uniquely associated with the individual slits. Therefore, although it is not true in the general case, there are specific experimental situations for which the quantum formalism does not predict any genuine third-order interference effects.  

It should be emphasized that even in the two-slit case there are exotic paths that pass through both slits and are contained in the second-order interference contributions. This means that a non-zero Sorkin parameter only accounts for complex paths that interact with all three slits. If they usually constitute a minimal correction \cite{quach2017}, they become more relevant as the operating wavelength increases with respect to the slit spacing. This is because the longer the wavelength, the greater the overlap between the wave functions of the individual slits, which gives more weight to the non-classical paths interacting with all the slits. This explains why the 2018 experiment by Govindaraj Rengaraj et al., that measured a non-zero Sorkin parameter, worked in the microwave regime \cite{rengaraj2018}. So, in quantum mechanics, when we are in frequency regimes (or experimental designs) where the exotic paths contributions are not relevant, there are no genuine third-order interference effects, but when we move from the physical to the psychological laboratory, things may change, and it is one of the aims of this article to investigate this possibility. 

To better understand what we mean when we talk about third-order contributions that would have the same relevance as second-order contributions, let us return to Hampton's experiment \cite{Hampton1988}. Imagine a new hypothetical situation (not considered by Hampton) where, in addition to question (a) about \emph{Fruit} and question (b) about \emph{Vegetable}, students are also asked to answer a question (c) about \emph{Flavoring}, i.e., to evaluate the membership and typicality of the available exemplars also with respect to the concept \emph{Flavoring}, and then the three questions associated with all possible two-word disjunctions: (d) \emph{Fruit or vegetable}, (e) \emph{Fruit or flavoring}, and (f) \emph{Vegetable or flavoring}, plus of course the three-word disjunction (g) \emph{Fruit or vegetable or flavoring}. In other words, in this hypothetical experiment, students are now asked seven different questions, allowing for the calculation of seven different probability functions and the associated Sorkin parameter (\ref{sorkin-uniform}). Now, when choosing a typical exemplar of the three-concept combination \emph{Fruit or vegetable or flavoring}, if students only consider the emergent meanings that result from considering at most two concepts at a time in their reasoning, and never the meaning associated with the full disjunction of the three concepts, the experimental data would not contain any genuine third-order interference contribution and one expect to find a Sorkin parameter close to zero. 

We are not aware of any tests for possible violations of the nullity of the Sorkin parameter in experiments with combinations of three concepts, and it is certainly our intention, in the near future, to carry out such an experiment. However, we can certainly reason as follows. The three two-word disjunctions, \emph{Fruit or vegetable}, \emph{Fruit or flavoring} and \emph{Vegetable or flavoring}, express different situations of doubt about the classification of a given exemplar of \emph{Food}. Imagine that an exemplar cannot be conveniently classified as either \emph{Fruit}, \emph{Vegetable}, or \emph{Flavoring}. This will lead the corresponding two-word conjunctions receiving higher scores, i.e., being overextended, precisely because they express a situation of doubt. However, the three-word conjunction, \emph{Fruit or vegetable or flavoring}, also conveys a meaning of doubt about how to classify the exemplar in question, and undoubtedly does so differently from the two-word conjunctions, since it explicitly contains all three words. The question that naturally arises is whether this three-word (three-slit-like) expression of doubt allows, in real experiments, meanings to emerge that are in no way deducible from two-word (two-slit-like) expressions of doubt. If this is the case, the equality $\epsilon(x|123)=0$ is going to be violated. But will it be a weak violation, as is usually the case in quantum mechanics, or a strong one? The cognitive test we have conducted, which is described in the next section and simulates single-slit, double-slit and triple-slit experiments, asking participants to place impact points directly on the equivalent of a detection screen, will allow us to offer an initial response to this question.

\section{The cognitive triple-slit experiment\label{doubleslitcognitive}}

In this section, we explain how our cognitive experiment was conducted in practice. For simplicity, we only considered a one-dimensional detection screen. Note that the typical interference fringes are a consequence of the slits' geometry, in the sense that it is their height, and the fact that they are parallel to each other, that are responsible for the fringe appearance, while it is their  distance from each other, and width, that are responsible for the alternation between high and low impact zones and diffraction phenomena. For the purposes of our experiment, however, it was not necessary to investigate the vertical dimension of the slits. 

\begin{figure}[htbp]
\begin{center}
\includegraphics[width=8cm]{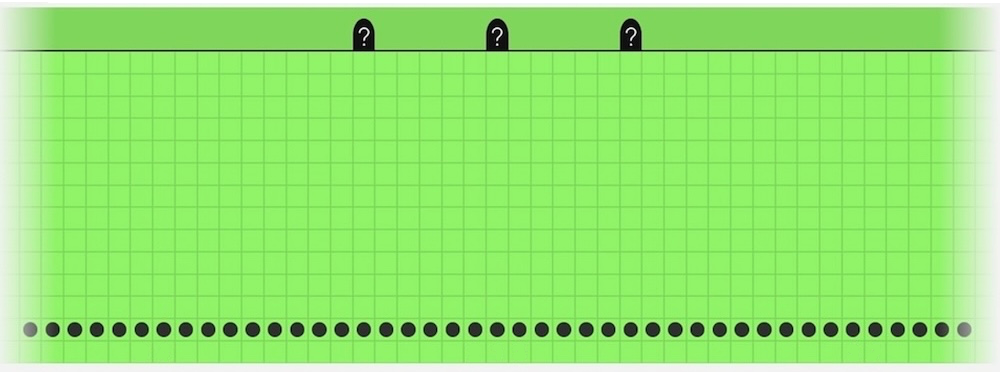}
\caption{(color online) The large rectangular area that was presented to the participants, with the row of $43$ identical portions of food, here in the situation with three openings.}
\label{largearea}
\end{center}
\end{figure}

More precisely, our detection screen consisted of $n=8d+3$ elementary cells of unit length, where $d=5$ corresponds to the distance between the slits (the number of cells separating them) in the slit plate. The latter was placed at a distance of $L=12$ cells from the detection screen, which was formed of $n=43$ cells, with discrete positions $x=1,\dots,n$. The slits were chosen to be of unit width and placed in correspondence with cells with numbers $n=16, 22, 28$ (see Figures~\ref{largearea} and \ref{Figure-p1p2p3+p12p23p13}).

A number $N=213$ of subjects participated in our interactive online experience in which they were asked to complete seven tasks in response to specific questions (corresponding to three one-slit, three double-slit and one triple-slit experiments). They were chosen at random from colleagues and friends and also recruited via a YouTube video on one of the authors' channel. They were all exposed to the same questions and experimental conditions (repeated measures design). Also, the order of the one-slit and two-slit experiments were randomized for each subject, to minimize order effects in data collection. The number of participants, recruited from March 6 to April 6, 2025, was actually $221$, but we excluded those who reported that the instructions were too difficult to understand or who took too much time (more than $30$ minutes) or too little time (less than $3$ minutes) to complete the experiment (the average time was $9.27$ minutes). 

More precisely, after agreeing to participate, each participant had to read an introductory text explaining the context of the experiment. Note that the introductory text, and subsequent explanatory texts, did not mention that it was a cognitive simulation of a quantum physics experiment, so as not to condition their choices on the basis of possible prior knowledge of the double-slit and possibly triple-slit experiments. For this reason, the different situations were presented to them in purely imaginative terms, as a description of an animal looking for food to eat (see below). In addition to the introductory text, each of the tasks was preceded by an explanation of the criterion by which they had to select a certain number of points (cells) on the detection screen. Basically, in each task participants had to select $7$ cells in sequence. The first selection corresponded to what was considered the best cell to answer the question asked, receiving (during analysis) a score of $7$ points. The second selected cell, which could not be the same because once a cell was selected the system disabled it and could not be selected a second time, received a score of $6$ points, and so on, until the seventh selected cell, which received only $1$ point. In other words, respondents answered the questions using a rank-ordering procedure, with scores from $7$ to $1$ assigned during analysis. This is of course not exactly what happens in a physics laboratory, where the measuring instrument records only one impact at a time. If each subject had been asked to select only one cell per task, however, the number of data collected might have been insufficient. Nevertheless, the patterns highlighted in this work remain identifiable regardless of the score that is attributed to the sequential choices of the participants, as we will explain in Section~\ref{analysisdata}. 

For the single-slit experiments, participants were asked to select cells that were the best examples of impacts caused by particles emerging from the slit in question. For the experiments with two and three slits, they were asked to select cells that were the best examples of a situation of doubt about the slit from which the particle emerged. To best convey these situations, the various questions were posed in the form of a bet. The source of particles behind the slit plate was not described, as this would have complicated the understanding of the experiment and introduced considerations of the position of the source relative to the three slits. For the participants, each slit played the role of a potential source, with the constraint that only one at a time could emit a particle. More precisely, following some preliminary questions, participants had to read the following text (see Figure~\ref{largearea}): \emph{This test evaluates how you interpret situations involving varying degrees of uncertainty. You will be presented with different situations and for each one you will be asked some questions. Although a question may seem similar to the previous one, you must always remain focused and answer as if it were the first time.} The following description ensued: \emph{There is a large rectangular area. On one side there is a row of $43$ identical portions of food. On the opposite side is a wall with one or more openings. A hungry animal emerges from an opening and moves towards the line of food. Once at the line, the animal eats the portion of food in front of it and goes away. You have no information about the nature of the animal. You don't know if the animal can sense the food, and if it can, you don't know which sense it uses (sight, smell, etc.). This means that its path could even be exploratory or erratic.} 

Participants had then to execute their first task, answering the following question: \emph{In this situation, the animal can only come out of one opening. You have bet a lot of money that you can guess which portion of food the animal will eat. What is your choice? To select a portion of food, click on it.} Next, they were asked to provide their best second choice, and so on, until the seventh. Then, the same situation was presented to them two more times, changing the place of the opening. Following these three single-opening (single-slit) experiments, participants had to move to their fourth task, which was different from the previous ones. 

The instruction for the fourth task was as follows: \emph{Many things have changed in this new situation: there are now two openings from which the animals can come out; now is you who decide which portions of food were eaten by the animals; each time a portion of food has been eaten, and the animal has left, another person, your opponent, will observe the situation and try to guess which opening the animal came from, taking into account which portion of food has been eaten; you have bet a lot of money that your opponent will not guess correctly; which portion of food should you choose for the animal to eat, so that your opponent cannot guess which opening it came from, and you win a lot of money? To select a portion of food, click on it. Note: Your opponent understands that there is uncertainty as to how the animal senses and reaches the food. In other words, in his/her reasoning, your opponent takes into account the fact that the animal may follow exploratory or irregular paths.}\footnote{Following Aerts’ conceptuality interpretation of the double-slit experiment (see Section~\ref{intro}), our experiment was designed to prompt participants to actualize outcomes in line with the idea that a detector selects the impact points most representative of the uncertainty inherent in the experimental context. This accounts for the apparent difference between the one-slit and double-slit situations, with an opponent introduced only in the latter. The opponent functioned merely as a stratagem to compel participants to focus on doubt, i.e., to select those cells that are most representative of the uncertainty about the opening from which the animal emerged. In the single-slit case, no opponent is needed, since the animal necessarily emerges from the only available opening.} Again, participants had to provide their best seven choices in order, and the situation was then repeated twice with the two openings relocated.

Following these three double-opening (double-slit) experiments, participants had to move to their seventh and last task. The instruction for it was as follows: \emph{This is the last variation. There are now three openings for the animals to come out of. Which portions of food should you choose for the animals to eat, so that your opponents (different each time) cannot guess which opening they came from, and you win a lot of money? As you did before, answer the question seven times, choosing the portions of food that are your first, second, third, fourth, fifth, sixth and seventh best choices. Note: Think carefully, now that there are three openings, the situation is more complex.}

\begin{figure}[htbp]
\begin{center}
\includegraphics[width=9cm]{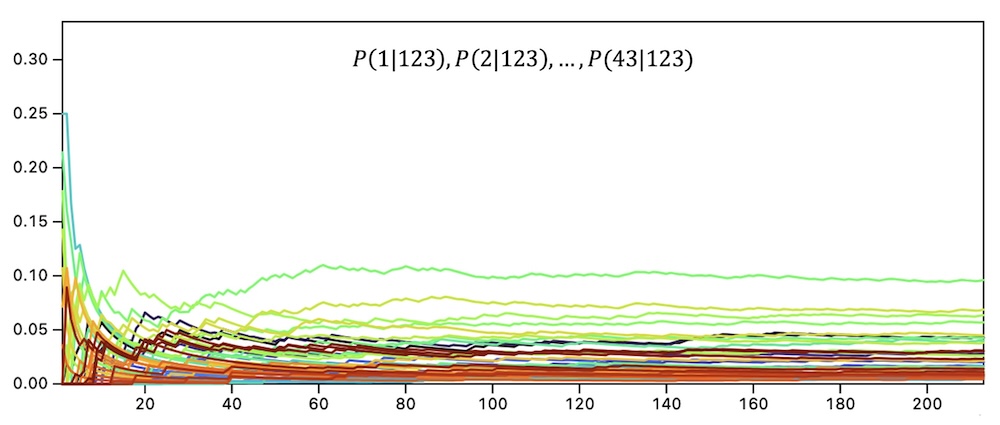}
\caption{(color online) The evolution of the three-slit probabilities $P(x|123)$, $x=1,\dots,43$, as a function of the number of participants who progressively took part in the cognitive experiment, from $N=1$ to $N=213$. We observe that the curves stabilize as the number of participants increases, consistent with the law of large numbers.}
\label{evolution}
\end{center}
\end{figure}

Upon completion of the cognitive test, each of the $N=213$ admitted participants distributed $\sum_{i=1}^7i=28$ points to $7$ different cells for each of the seven tasks. Thus, a total of $N_{\rm tot}=28N = 5964$ datapoints were assigned to the $n=43$ cells of the detection screen. If $N(x|1)$, $N(x|2)$ and $N(x|3)$ denote the number of points assigned to cell $x$, when only slit-$1$, slit-$2$ and slit-$3$ are open, respectively,  dividing them by  $N_{\rm tot}$ one obtains the corresponding one-slit experimental probabilities $P(x|1)$, $P(x|2)$ and $P(x|3)$. One does the same with the numbers $N(x|12)$, $N(x|23)$ and $N(x|13)$, when only `slit-$1$ and $2$', `slit-$2$ and $3$', and `slit-$1$ and $3$' are open, respectively, and with the number $N(x|123)$, when all three slits are open, obtaining in this way the two-slit experimental probabilities $P(x|12)$, $P(x|23)$ and $P(x|13)$, and the three-slit probability $P(x|123)$, $x=1,\dots,n$. All these probabilities are reported in Table~\ref{table1}, with the Sorkin parameter $\epsilon(x|123)$ and its normalized version $\kappa(x|123)= P_{\rm max}^{-1}\, \epsilon(x|123)$, $P_{\rm max}= \sup_{x}P(x|123)$. Note that assuming a binomial distribution, the maximum value for the standard error is ${\rm SE}(19)\approx 0.00382$, where $x=19$ corresponds to the maximum probability $P_{\rm max}=P(19|123)\approx 0.0961$. If we consider a confidence level of $95\%$, the $z$-value is $1.96$ and the confidence interval becomes:  $P(19|123) \approx 0.0961\pm 1.96\cdot 0.00382=0.0961\pm 0.0075$; see also Figure~\ref{evolution}. 

\begin{scriptsize}
\begin{center}
\begin{table}
\begin{tabular}{|c c c c c c c c c c|} 
\hline 
 $\!\!x\!\!$ & $\!\!P(x|1)\!\!$ & $\!\!P(x|2)\!\!$ & $\!\!P(x|3)\!\!$ & $\!\!P(x|12)\!\!$ & $\!\!P(x|23)\!\!$ & $\!\!P(x|13)\!\!$ & $\!\!P(x|123)\!\!$ & $\!\!\epsilon(x|123)\!\!$ & $\!\!\kappa(x|123)\!\!$\\
 \hline\hline
  1 & 0.013 & 0.012 & 0.011 & 0.027 & 0.055 & 0.044 & 0.043 & -0.030 & -0.309\\ 
  
2 & 0.012 & 0.006 & 0.008 & 0.032 & 0.049 & 0.042 & 0.043 & -0.031 & -0.319\\ 

3 & 0.011 & 0.005 & 0.007 & 0.024 & 0.045 & 0.023 & 0.028 & -0.026 & -0.276\\ 

4 & 0.014 & 0.007 & 0.007 & 0.021 & 0.037 & 0.021 & 0.020 & -0.024 & -0.248\\ 

5 & 0.012 & 0.005 & 0.005 & 0.010 & 0.027 & 0.011 & 0.017 & -0.008 & -0.080\\ 

6 & 0.013 & 0.006 & 0.005 & 0.010 & 0.021 & 0.019 & 0.020 & -0.004 & -0.046\\ 

7 & 0.011 & 0.009 & 0.004 & 0.014 & 0.016 & 0.012 & 0.009 & -0.011 & -0.114\\ 

8 & 0.015 & 0.006 & 0.005 & 0.010 & 0.018 & 0.008 & 0.006 & -0.010 & -0.102\\ 

9 & 0.015 & 0.007 & 0.005 & 0.011 & 0.009 & 0.013 & 0.008 & -0.004 & -0.044\\ 

10 & 0.021 & 0.008 & 0.002 & 0.012 & 0.011 & 0.005 & 0.007 & -0.001 & -0.005\\ 

11 & 0.019 & 0.013 & 0.004 & 0.013 & 0.004 & 0.014 & 0.008 & -0.001 & -0.005\\ 

12 & 0.029 & 0.013 & 0.002 & 0.012 & 0.008 & 0.006 & 0.005 & 0.001 & 0.013\\ 

13 & 0.052 & 0.008 & 0.006 & 0.010 & 0.009 & 0.009 & 0.017 & 0.020 & 0.206\\ 

14 & 0.077 & 0.012 & 0.006 & 0.012 & 0.008 & 0.008 & 0.010 & 0.023 & 0.237\\ 

15 & 0.109 & 0.013 & 0.007 & 0.021 & 0.005 & 0.019 & 0.013 & 0.026 & 0.269\\ 

16 & 0.175 & 0.023 & 0.008 & 0.037 & 0.008 & 0.039 & 0.040 & 0.053 & 0.549\\ 

17 & 0.106 & 0.016 & 0.004 & 0.047 & 0.009 & 0.019 & 0.015 & 0.007 & 0.076\\ 

18 & 0.064 & 0.022 & 0.007 & 0.074 & 0.008 & 0.014 & 0.042 & 0.009 & 0.098\\ 

19 & 0.052 & 0.057 & 0.009 & 0.117 & 0.015 & 0.022 & 0.096 & 0.033 & 0.340\\ 

20 & 0.020 & 0.090 & 0.009 & 0.063 & 0.017 & 0.048 & 0.057 & 0.011 & 0.116\\ 

21 & 0.019 & 0.120 & 0.014 & 0.041 & 0.018 & 0.083 & 0.030 & -0.013 & -0.140\\ 

22 & 0.017 & 0.182 & 0.017 & 0.032 & 0.049 & 0.133 & 0.064 & -0.007 & -0.069\\ 

23 & 0.015 & 0.114 & 0.022 & 0.014 & 0.043 & 0.078 & 0.026 & -0.013 & -0.134\\ 

24 & 0.012 & 0.066 & 0.026 & 0.009 & 0.076 & 0.042 & 0.038 & -0.012 & -0.121\\ 

25 & 0.011 & 0.048 & 0.053 & 0.009 & 0.117 & 0.021 & 0.069 & 0.008 & 0.088\\ 

26 & 0.009 & 0.021 & 0.081 & 0.011 & 0.065 & 0.006 & 0.045 & 0.028 & 0.289\\ 

27 & 0.008 & 0.013 & 0.120 & 0.008 & 0.037 & 0.010 & 0.010 & 0.021 & 0.218\\ 

28 & 0.005 & 0.008 & 0.172 & 0.010 & 0.037 & 0.027 & 0.021 & 0.034 & 0.354\\ 

29 & 0.004 & 0.014 & 0.111 & 0.009 & 0.013 & 0.015 & 0.011 & 0.029 & 0.306\\ 

30 & 0.006 & 0.005 & 0.073 & 0.009 & 0.007 & 0.006 & 0.007 & 0.021 & 0.215\\ 

31 & 0.003 & 0.008 & 0.045 & 0.008 & 0.007 & 0.006 & 0.008 & 0.013 & 0.136\\ 

32 & 0.002 & 0.007 & 0.017 & 0.014 & 0.008 & 0.009 & 0.014 & 0.002 & 0.024\\ 

33 & 0.003 & 0.007 & 0.017 & 0.006 & 0.007 & 0.002 & 0.008 & 0.007 & 0.069\\ 

34 & 0.003 & 0.005 & 0.014 & 0.013 & 0.008 & 0.007 & 0.007 & -0.005 & -0.048\\ 

35 & 0.002 & 0.005 & 0.014 & 0.011 & 0.009 & 0.004 & 0.004 & -0.005 & -0.052\\ 

36 & 0.002 & 0.004 & 0.009 & 0.013 & 0.010 & 0.010 & 0.007 & -0.011 & -0.116\\ 

37 & 0.005 & 0.006 & 0.008 & 0.016 & 0.008 & 0.015 & 0.010 & -0.011 & -0.109\\ 

38 & 0.005 & 0.005 & 0.018 & 0.012 & 0.011 & 0.009 & 0.008 & -0.004 & -0.040\\ 

39 & 0.002 & 0.003 & 0.005 & 0.026 & 0.013 & 0.011 & 0.011 & -0.019 & -0.202\\ 

40 & 0.004 & 0.004 & 0.008 & 0.026 & 0.012 & 0.012 & 0.014 & -0.014 & -0.148\\ 

41 & 0.005 & 0.005 & 0.013 & 0.032 & 0.017 & 0.025 & 0.023 & -0.018 & -0.191\\ 

42 & 0.008 & 0.003 & 0.005 & 0.049 & 0.022 & 0.034 & 0.030 & -0.035 & -0.361\\ 

43 & 0.010 & 0.009 & 0.017 & 0.043 & 0.030 & 0.038 & 0.031 & -0.031 & -0.323\\ 
 \hline
\end{tabular}
\caption{The experimental probabilities for the $43$ cell and the Sorkin parameters $\epsilon(x|123)$ and $\kappa(x|123)= P_{\rm max}^{-1}\, \epsilon(x|123)$, $P_{\rm max}= \sup_{x}P(x|123)$.}
\label{table1}
\end{table}
\end{center}
\end{scriptsize}

\section{Analysis of the experimental data\label{analysisdata}}

In this section, we analyze the data from our experiments and compare them with those typically obtained in physics laboratories. We begin by summarizing some of the values of the Sorkin parameter reported in the literature. Note that to allow for an easier comparison, normalized versions of the parameter are usually considered. In \cite{sinhaetal2010}, a normalization factor $\delta(x|123)= |I(x|12)| + |I(x|23)| + |I(x|13)|$ was used. Performing a three-slit experiment with photons, $\epsilon'(x|123)\equiv \delta^{-1}(x|123)\epsilon(x|123)$ was measured at the central maximum of the triple-slit interference pattern. The following values were obtained: $\epsilon'(x|123) = 0.0073 \pm 0.0018$, when using a laser source and a power meter for detection; $\epsilon'(x|123) = 0.0034 \pm 0.0038$, when attenuating the laser to single-photon level and a silicon avalanche photo-diode for detection; $\epsilon'(x|123) = 0.0064 \pm 0.0119$, when using the heralded single-photon source. Hence, overall, they found $\epsilon'(x|123)$ to be between $-0.0055$ and $0.0184$. In \cite{sollneretal2012},  an experiment with a three-path interferometer was performed and for the intensity maximum of the triple-slit interference pattern the following values were found: $\epsilon'(x|123)=-0.0015 \pm 0.0029$; $\epsilon'(x|123)=-0.0050 \pm 0.0018$; $\epsilon'(x|123)=-0.0065 \pm 0.00198$; $\epsilon'(x|123)=-0.0142 \pm 0.0010$, for increasing intensities at the three-path maximum. Thus, overall, the authors found $\epsilon'(x|123)$ to be between $-0.0152$ and $0.0014$. 

In \cite{sinhaetal2014}, it was proposed to use the intensity at the central maximum  as a normalization factor: $\kappa(x|123)\equiv P_{\rm max}^{-1}\, \epsilon(x|123)$, $P_{\rm max}= \sup_{x}P(x|123)$.\footnote{Note that $\epsilon(x|123)$ can range from $-2$ to $+2$. Since the minimal value of $P_{\rm max}$ is $1/n$, $\kappa(x|123)$  can range from $-2n$ to $+2n$. These are of course purely algebraic bounds.} Using the Feynman path integral formalism in the thin-slit approximation, it was shown that  $\kappa(x|123)$ is strongly dependent on certain experimental parameters and increases with an increase in wavelength. For the parameters used in \cite{sinhaetal2010}, $\kappa(x|123)$ at the central maximum was found to be of the order of $10^{-6}$. However, for an incident beam of wavelength $4 \, {\rm cm}$ (microwave regime), slit width of $120 \, {\rm cm}$ and inter-slit distance of $400\, {\rm cm}$, their theoretical estimate of  $\kappa(x|123)$ was of the order of $10^{-3}$. In \cite{sinhatal2015}, an upper limit for $|\kappa(123)|$ was derived, in the regime where the slit width $w$ and the inter-slit distance $d$ are such that $d\gg w$, and $w\gg {\lambda\over 2\pi}$, with $\lambda$ the wavelength: $|\kappa_{\rm max}(123)|\approx 0.03 \lambda^{3\over 2}d^{-{1\over 2}}w^{-1}$. 

In \cite{maga2016}, it was shown that looped trajectories of photons are due to the near-field component of the wavefunction, which leads to an interaction among the three slits. Hence, by modifying the dimensions of the three slit structure or by changing the wavelength of the optical excitation, the authors were able to increase the probability of occurrence of these trajectories, obtaining for instance values of $\kappa(x|123)$, for the central maximum, ranging from $-0.35$ to $0.15$. In their experiment they used a classical light from a tunable diode laser with wavelengths from $750\, {\rm nm}$ to $790\, {\rm nm}$, slit width $w=215\, {\rm nm}$, and inter-slit distance $d=4.6\, \mu{\rm m}$. In \cite{rengaraj2018}, experiments were performed in the microwaves domain, using pyramidal horn antennas as sources. A source-detector separation of $2.5\, {\rm m}$ was used ($1.25\, {\rm m}$ between source and slits plate and same distance between the slit plate and the detection screen). Non-zero values of $\kappa(x|123)$ of the order of $10^{-2}$ were measured, much above the error bound.  

\begin{figure}[htbp]
\begin{center}
\includegraphics[width=13cm]{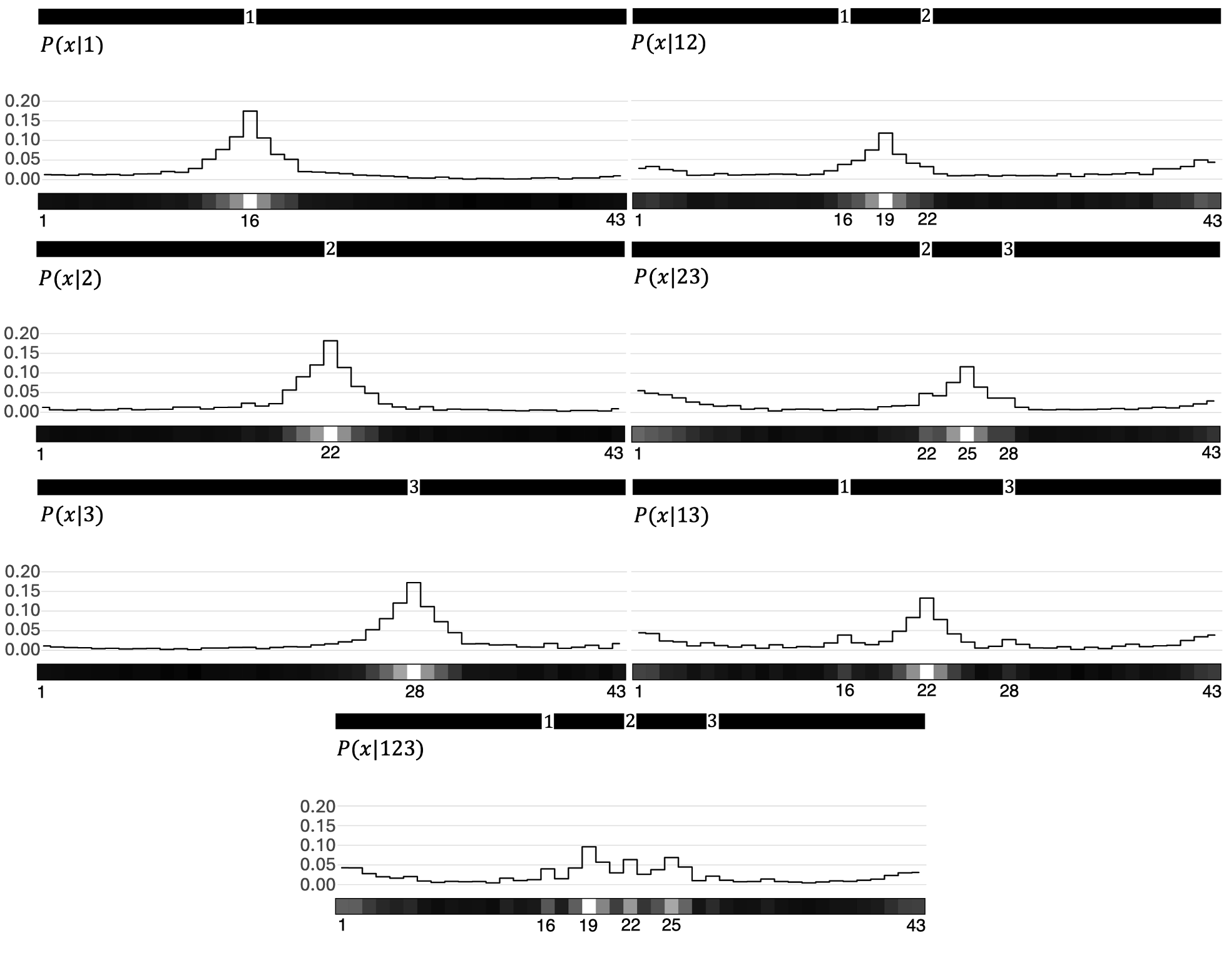}
\caption{Graphic representation of the data reported in Table~\ref{table1}. Left column: Plots of the one-slit probabilities $P(x|1)$, $P(x|2)$ and $P(x|3)$. Right column: Plots of the two-slit probabilities $P(x|12)$, $P(x|23)$ and $P(x|13)$. Bottom center figure: Plot of the three-slit probability $P(x|123)$.}
\label{Figure-p1p2p3+p12p23p13}
\end{center}
\end{figure}

Having summarized the experimental values reported in the physics literature, we can proceed to analyze our cognitive data, which we have represented graphically in Figures~\ref{Figure-p1p2p3+p12p23p13} and \ref{sorkin}. Those for the three one-slit experiments, described in the left column of Figure~\ref{Figure-p1p2p3+p12p23p13}, do not present any problems of interpretation. The question posed to the participants did not represent a situation of doubt, as there was certainty about the slit (the opening) from which the quantum entity (the animal) would emerge, and this explains the single peak forming in front of each of the slits.

Considering the data for the experiments with two slits, described in the right column of Figure~\ref{Figure-p1p2p3+p12p23p13}, we observe that all three experiments show a clear central peak. This was of course to be expected, as the question posed attempted to maximize doubt about the slit (the opening) from which the quantum entity (the animal) emerged, and a central impact does indeed reflect maximum uncertainty.  Another element common to the three two-slit probabilities in Figure~\ref{Figure-p1p2p3+p12p23p13} is the fact that participants also placed many of their points at the extremes of the detection screen. The idea here is that by moving as far away as possible from the two slits, it becomes more difficult to discriminate between them, from a spatial perspective, so in some way a new situation of (local) maximum doubt is obtained about the slit from which the entity emerged. However, we also observe that the maximum is not always found at the outermost cells; this is the case, for example, in the graph for $P(x|12)$. One participant commented that choosing the most distant cells would have favored the closest slit. The logic behind this thought could be the following. If propagation from the slit to the point of impact has a cost proportional to the distance traveled, and resources are limited, then an entity emerging from slit-$1$ will be able to go further to the left than an identical entity (with same resources) emerging from slit-$2$. This  suggests that if we had used a larger screen, a wide secondary fringe could have possible formed, at a distance not necessarily corresponding to its extreme sides, and this could be one of the cognitive mechanism at the origin of secondary regions of probability overextension, in addition to the central one. 

There is another aspect that is particularly evident in the graph of $P(x|13)$, where one can observe two additional secondary peaks in correspondence with the positions of the slits. These can also be glimpsed in the graph of $P(x|23)$. A sufficient number of participants favored this choice to make it emerge in the structure. The positions corresponding to the slits appear to be attractive in some way. Is this a rational or irrational choice? There are two possible interpretations. One is to consider that this type of choice minimizes rather than maximizes doubt. This would mean that rational thought processes were not the only ones involved in the decision-making. The other interpretation is that, since the animal’s path could be exploratory or erratic, placing the food portion in correspondence with the opening (i.e., the slit) could be an equally valid way to deceive the opponent. Indeed, if we assume an erratic animal, then the reasoning that maximizes the distance between the food portion and the opening is no longer valid. In other words, strategies could consider doubt expressed not only at the level of the position of the food, but also at the level of its meaning in relation to the animal's exploratory behavior.

Turning now to the situation with three slits (the central graph at the bottom of Figure~\ref{Figure-p1p2p3+p12p23p13}), we always have accumulation points at the ends of the screen, and secondary peaks at each of the three slits, plus two major peaks at the midpoint between the slits. We also observe a strong asymmetry between left and right as far as the two midpoints are concerned, the midpoint between slits $1$ and $2$ being chosen much more frequently than the midpoint between slits $2$ and $3$. The left side of the screen was also favored for peaks in correspondence with slits $1$ and $2$ and for the accumulation points at the ends of the screen, confirming a general bias favoring the left side of the screen. The origin of this asymmetry remains unclear. It may be attributed to the fact that participants are from a left-to-right reading culture, or it could be due to the population's asymmetry between left-handed and right-handed individuals, but these are hypotheses that would require validation.

\begin{figure}[htbp]
\begin{center}
\includegraphics[width=8cm]{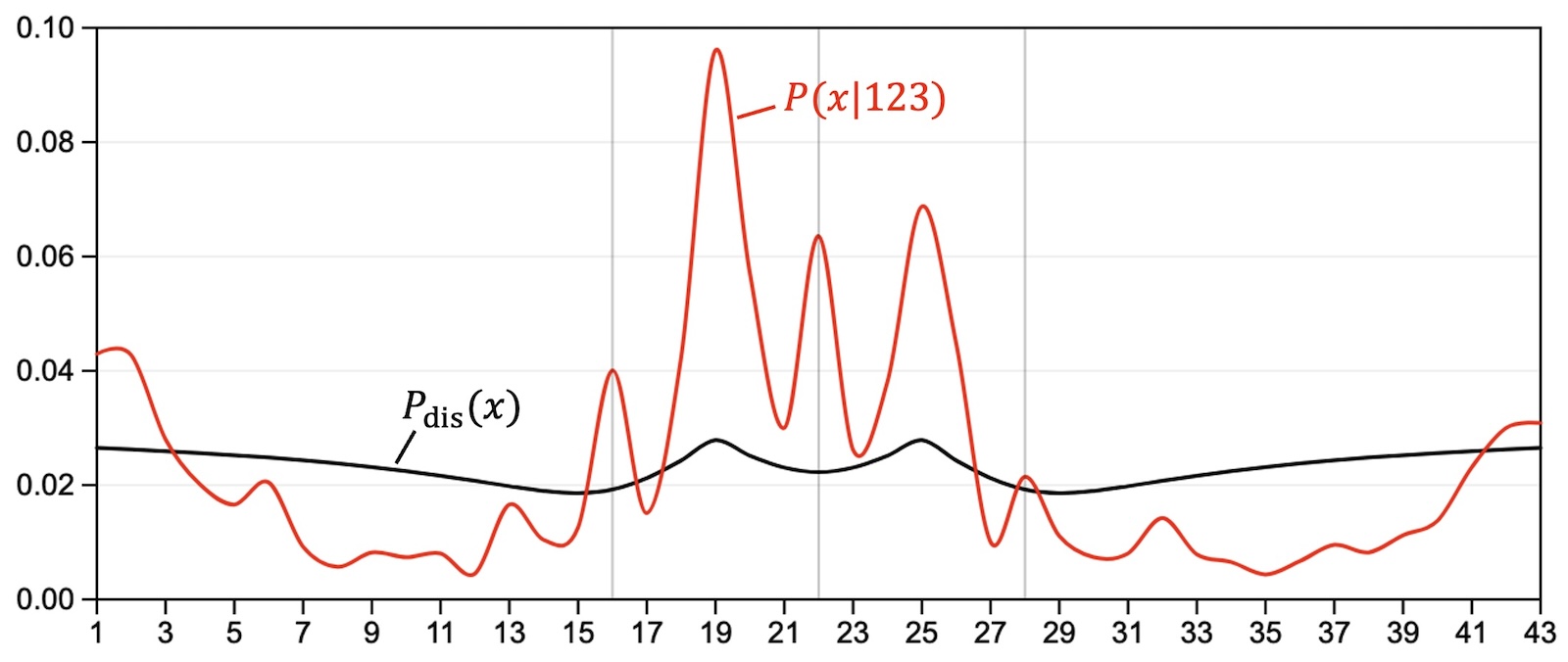}
\caption{(color online)  Plot of the triple-slit probability $P(x|123)$ (red color) together with the probability $P_{\rm dis}(x)$ (black color) describing a reasoning based on the evaluation of distances between slits and impact points. To make the variations of $P_{\rm dis}(x)$ easier to read, it has been drawn using a distance $L=3$ (instead of $L=12$). As $L$ decreases (near-field limit), $P_{\rm dis}(x)$ increases at the edges of the screen, exceeding its value at the center.}
\label{Figure-distances}
\end{center}
\end{figure}

When comparing our graph with those typically obtained in physics laboratories, the most striking difference is perhaps the fact that the central fringe is not the one of maximum intensity. A possible overall explanation for the existence of the two main peaks between slits $1$ and $2$, and between slits $2$ and $3$, for the secondary peak in correspondence with slit $3$, and for the two regions of accumulation of lesser intensity at the extremes of the screen, may be the following. When choosing a point $x$ on the screen, which should maximize the doubt about the slit from which the entity emerged, a participant will tend to associate with each slit a probability $\rho_i(x)$, $i=1,2,3$, inversely proportional to the distance $L_i$ between the slit and the point $x$ in question. In other words, the shorter the distance between the slit and point $x$, the more likely it is that the entity that produced that point emerged from that slit. More specifically, following Figure~\ref{a1}, we can write: 
\begin{eqnarray}
&&L_1=\sqrt{L^2+\left(x+d\right)^2},\quad L_2=\sqrt{L^2+x^2},\quad L_3=\sqrt{L^2+\left(x-d\right)^2} \nonumber\\
&&\rho_i(x)={A\over L_i},\quad A={L_1L_2L_3\over L_1L_2+L_2L_3+L_1L_3}
\label{rho}
\end{eqnarray}
Considering then the maximum $\rho(x)={\rm max} \{\rho_1(x), \rho_2(x), \rho_3(x)\}$, a participant trying to identify a point $x$ on the screen that maximizes the doubt about the slit from which the entity emerged will intuitively identify a position $x$ that minimizes $\rho(x)$. So, if  $\overline{\rho}(x)=1-\rho(x)$ and $B=\sum_x \overline{\rho}(x)$, we can define $P_{\rm dis}(x)=B^{-1}\overline{\rho}(x)$ as the probability for selecting a point $x$ when using a reasoning based on competing distances. In Figure~\ref{Figure-distances}, we have plotted $P_{\rm dis}(x)$ together with $P(x|123)$. We see that $P_{\rm dis}(x)$ has its two main peaks precisely between slits $1$ and $2$, and between slits $2$ and $3$. This would explain the participants' choice to favor those positions. Furthermore, the value of $P_{\rm dis}(x)$ is higher between the two slits than outside them, and this would explain the secondary choice of the central position. Finally, $P_{\rm dis}(x)$ shows accumulation points at the ends of the screen, which is also in line with the participants' choices. 

\begin{figure}[htbp]
\begin{center}
\includegraphics[width=13cm]{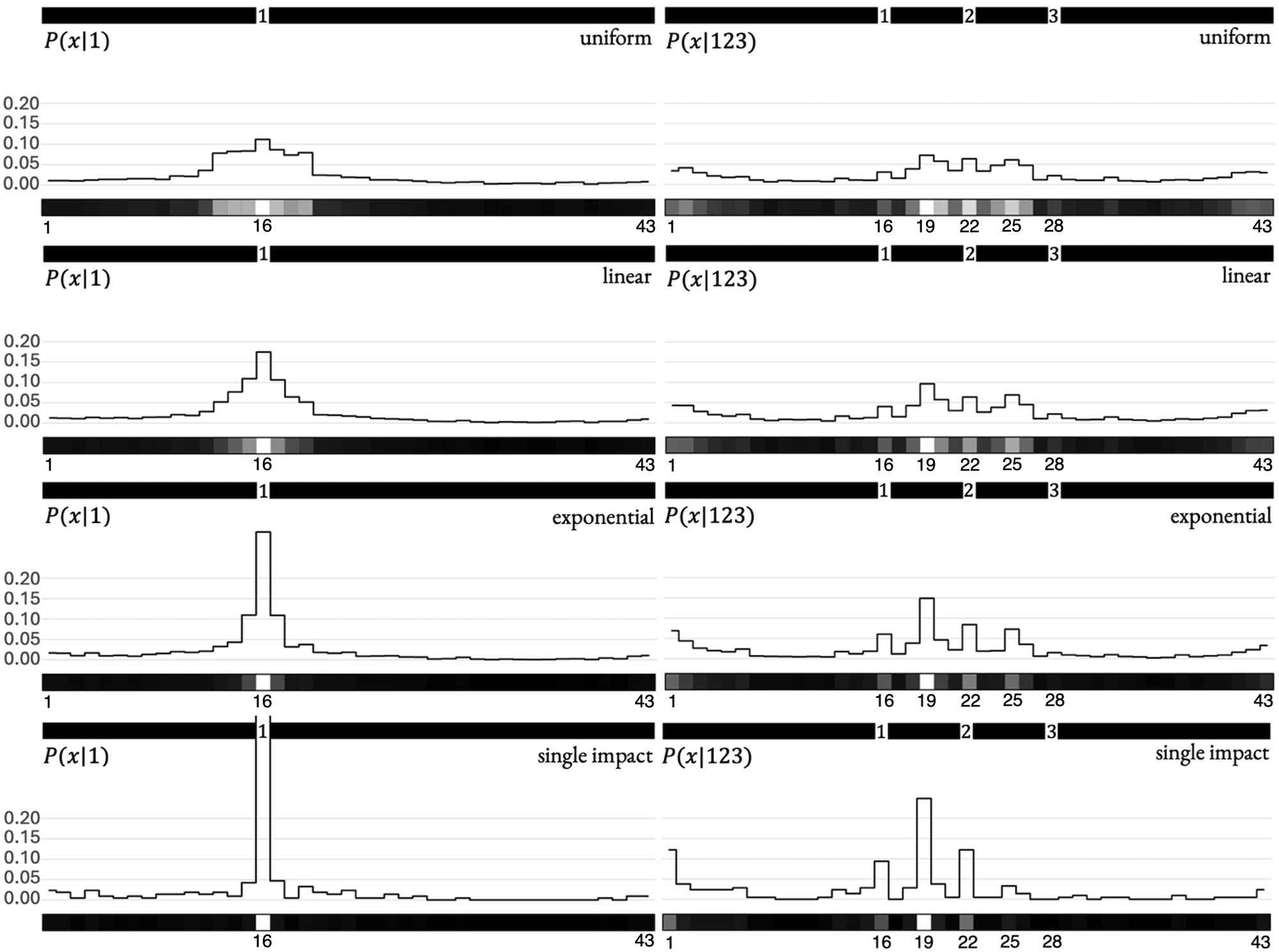}
\caption{A comparison of the data when changing the score assignment applied during analysis. In the left and right columns, the $P(x|1)$ and $P(x|123)$ probabilities are graphically represented, respectively. The first line corresponds to a uniform assignment (same score to all seven choices), the second line to a linearly decreasing assignment, the third line to an exponentially decreasing assignment, and finally the graphs of the last line only take into account participants' best (first) choice. We observe that, as more weight is given to the first choices, the graphs become more sharply peaked.}
\label{comparison}
\end{center}
\end{figure}

\begin{figure}[htbp]
\begin{center}
\includegraphics[width=8cm]{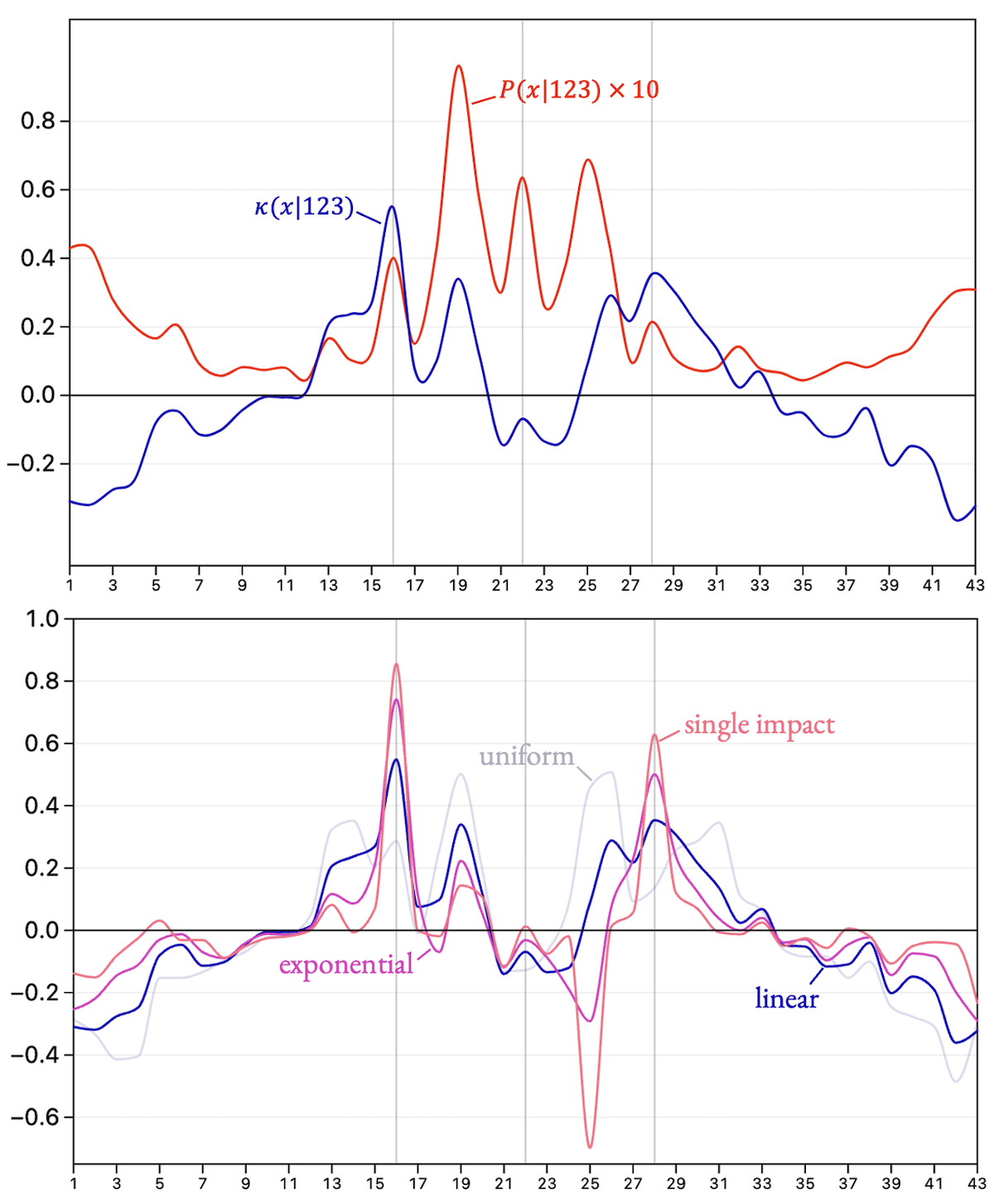}
\caption{(color online)  Top: Plot of the normalized Sorkin parameter $\kappa(x|123)$ (blue color). The probability $P(x|123)$, multiplied by a factor of $10$, is also plotted (red color), to give a clearer understanding of the modulation in the plot. A continuous approximation of the discrete functions was used to enhance interpretability in the visualization. Bottom: Plots of the normalized Sorkin parameter $\kappa(x|123)$ for different choices of the scores attributed to the participants' choices (see Figure~\ref{comparison}): uniform (light gray color), linear (blue color), exponential (magenta color) and single impact (pink color).}
\label{sorkin}
\end{center}
\end{figure}

It is interesting to ask whether it is possible to associate an effective wavelength that can account, at least approximately, for the observed patterns. For this, it is important to firstly note that the widths of our peaks depend on the scores attributed to the sequential choices of the participants. The data reported in Table~\ref{table1} correspond to the choice of a linear scale: the first choice receive $7$ points, the second $6$ points, and so on, with the seventh receiving $1$ point, for a total of $28$ points per participant. However, we could have chosen to assign the same score to all seven sequential choices (uniform scale), in which case all seven choices receive $1$ point, for a total of $7$ points per participant. Another possibility is to use an exponential scale, with the first choice receiving $64$ points, the second $32$ points, and so on, with the seventh receiving $1$ point, for a total of $127$ points per participant. Finally, it is possible to consider the situation closest to what happens in a physics laboratory, using only the first choice of each participant, hence $1$ point per participant. In Figure~\ref{comparison}, data relative to these four different choices of scales for the score assignment are compared, for the single-slit probabilities $P(x|1)$ and the three-slit probabilities $P(x|123)$.  

To estimate the value of an effective wavelength $\lambda$ involved in the formation of the single-slit peaks, we have to use the graphs where only single answers are considered, otherwise the width of the peaks arbitrarily depends on the score assignment applied during analysis. We see in Figure~\ref{comparison} (fourth line) that peaks have a width of approximately $\Delta x\approx 3$. If we use the Fraunhofer diffraction formula (see Appendix~\ref{appendixa}) $\Delta x \approx 2{\lambda L\over a}$, where $a$ is the width of the slit, since we have $a=1$ and $L=12$, the effective wavelength is $\lambda \approx a {\Delta x \over 2L}\approx {3\over 24}\approx 0.125$. 
On the other hand, If we apply the formula $\lambda \approx{xd\over L}$ (see Appendix~\ref{appendixa}), giving the approximate distance $x$ at which a second maximum (the first being that of the central fringe) for the probability forms in the far field approximation,\footnote{Note however that, as we will discuss later in the article, this approximation does not adequately describe our experiment, given that the nullity of Sorkin's parameter is strongly violated.} considering $d=6$ and $x\approx {d\over 2}=3$ (see Figure~\ref{Figure-p1p2p3+p12p23p13}),  we find an effective wavelength $\lambda\approx {d^2\over 2L}=1.5$, which is clearly not compatible with the previously determined $\lambda\approx 0.125$ describing an effective diffraction mechanism. This means that there is no way to associate a single effective wavelength to our cognitive experiments, and this would be an important difference with the situation in physics laboratories (see also the discussion in Section~\ref{conclusion}).

Let us now consider the normalized Sorkin parameter $\kappa(x|123)$. Its maximum value is for $x=16$, i.e., in correspondence with slit-$1$: $\kappa(16|123)=0.549$; see Table~\ref{table1}, and the right graph of Figure~\ref{sorkin}. In the left graph of Figure~\ref{sorkin}, we have also plotted $\kappa(x|123)$ for the different score assignments (see also Figure~\ref{comparison}). The maximum values obtained, for the normalized Sorkin parameter, are as follows: $0.508$ (uniform), $0.549$ (linear), $0.741$ (exponential) and $0.855$ (single impact). 

The values obtained are much higher than those typically obtained in physics laboratories. This reveals the presence of significant deviations of the probability $P(x|123)$ both from the classical situation and from the idealized quantum situation, when the principle of superposition can be applied naively. In other words, according to our quantum modeling of the experiment, in their decision-making participants were also guided by considerations that led to an overall assessment of the situation, taking into account the three slits together in their reasoning, rather than two at a time. This generated significant third-order interference-like contributions, which in physics can only be obtained in near-field situations, i.e., when the detection screen is not located far from the slits and/or the slits are very close together (distances on the order of the wavelength or less), so that non-radiative modes, such as evanescent waves, can easily mediate coupling between the slits and the detector, and/or between the slits.

It could be said that since the human cognitive instrument apparently uses different effective wavelengths at the same time, and wavelengths determine the scale of near-field dominance, in our double and triple-slit cognitive experiments we are always in regimes where near-field effects are important, hence the strong violation of the nullity of Sorkin parameter.

\section{Concluding remarks\label{conclusion}}

In our cognitive experiment, we observed some interesting similarities with physics experiments. For example, it was not obvious that additional fringes would form, in addition to those between the two slits. This means that there is enough richness in human cognitive processes to produce complex patterns that resemble those produced by interference phenomena in quantum physical processes. However, some important differences were also observed, which were nevertheless expected.

Our design was a sort of hybrid between a purely cognitive experiment, such as the one conducted by Hampton \cite{Hampton1988}, described in Section~\ref{intro}, and a physics experiment. Our motivation in proceeding in this way was to test one of the explanations offered by the conceptuality interpretation, consisting in explaining the interference fringes, in particular the central one, as regions of actualization of a situation of doubt about the slit from which the quantum entity emerged. 

It is important to note that although the strong variability of a fringe-like pattern can be interpreted as the effect of an interference phenomenon, there are no actual waves from which such pattern would originate in a cognitive experiment. Quantum wave functions should not be interpreted as real waves either, being clear that they exist in a configuration space that is distinct from the physical space, whose number of dimensions grows proportionally to the number of entities present. According to the conceptuality interpretation, what we interpret in quantum mechanics as interference phenomena, associated with a wave-like nature of the entities under study, would more fundamentally be the result of a variability in the meanings associated with the different outcomes, when these are organized and represented according to a certain logic. In other words, by recognizing that there is a more fundamental level governed by meaning and cognitive processes, it would be possible to overcome the wave-particle duality and understand the formation of patterns whose origin would not be wave-like but genuinely cognitive-like.

One possible objection to the conceptuality interpretation, in relation to its interpretation of the double-slit and triple-slit experiments, is that the patterns observed in physics laboratories are usually very regular, whereas those obtained by analyzing cognitive experiments \`{a} la Hampton, or like ours, present much more variability. It is not difficult to respond to this type of criticism. If we truly live in a \emph{pancognitivist} reality, as the conceptuality interpretation suggests, then the structures of meaning that have emerged within our human culture represent a very recent episode and have not yet had the opportunity to stabilize and organize themselves in the same way as the much older cognitive structures that supposedly form our physical universe. 

In the case of human cognition, things are also complicated by the fact that there are at least two lines to go from abstract to concrete concepts, while for entities in the material world only one of these lines would be active (see \cite{aertsetal2020,aertssassoli2024} for an in-depth discussion of this point). Cognitive experiments should therefore only be considered qualitatively similar to experiments in physics laboratories, without expecting them to correspond to the latter in every possible aspect. In any case, even between comparable cognitive experiments, significant structural differences are to be expected, as is the case between ours and Hampton's experiments, where for example in one there is a relatively intense central peak between the two slits, whereas in the other such a peak is absent (see Figure~\ref{Figure-FV-onedimensional}).

In conclusion, our cognitive triple-slit experiment adds to our understanding of human cognitive processes in situations of uncertainty and provides further confirmation of the explanatory power contained in the conceptuality interpretation. The observed patterns, structurally reminiscent of those found in quantum interference experiments, speak to a deeper interpretative layer, one where meaning and information play the role traditionally attributed to physical interactions. In this sense, our results not only support the idea that cognition can manifest interference-like effects, but also reinforce the plausibility of a pancognitivist reality, in which physical and human cognitive processes would be governed by analogous principles operating at different levels of abstraction.

\begin{appendices}                       

\section{The interference fringes in the double and triple-slit experiment\label{appendixa}}

\begin{figure}[htbp]
\begin{center}
\includegraphics[width=13cm]{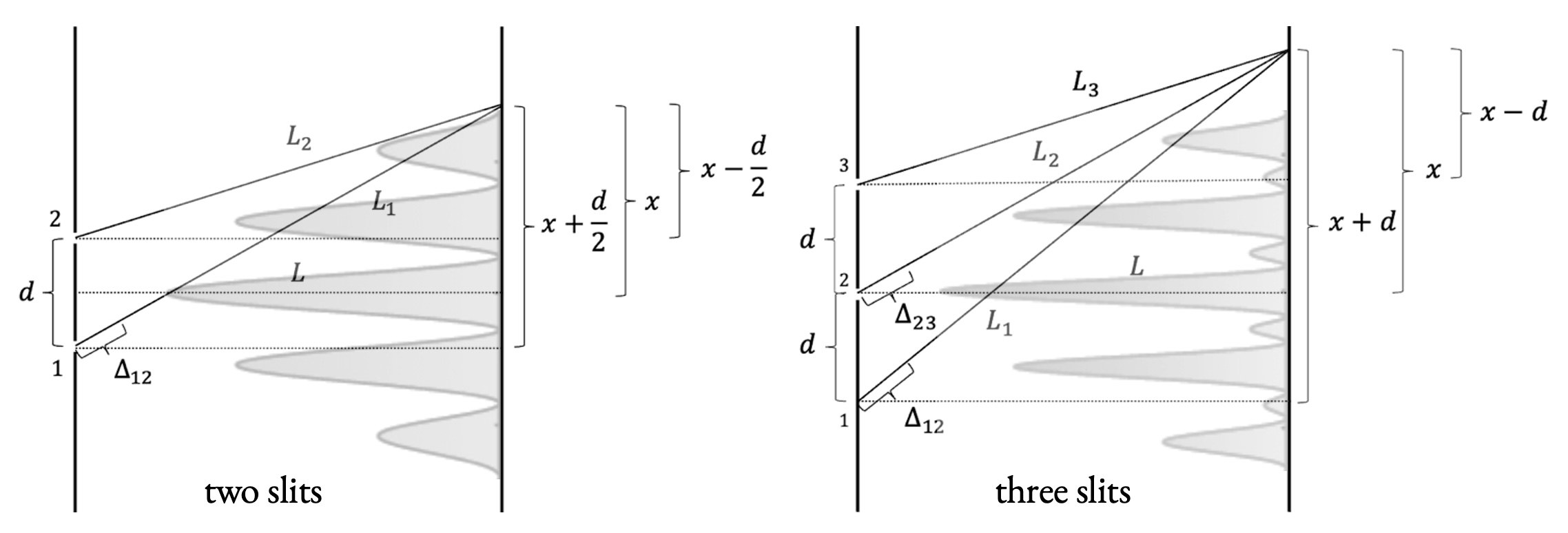}
\caption{Path lengths from the slits to a screen point $x$, and their differences, for the two-slit (left) and three-slit (right) cases.}
\label{a1}
\end{center}
\end{figure}

In this appendix we derive the positions of interference fringes in double- and triple-slit experiments, and the typical peak width in the single-slit case, within the Fraunhofer regime. We begin with the two-slit setup, initially neglecting slit width. Let $L$ and $d$ be the distances between the detector screen and the slit plate, and between the two slits, respectively; see Figure~\ref{a1}. 
Clearly, $L_1^2=L^2+\left(x+{d\over 2}\right)^2$ and $L_2^2=L^2+\left(x-{d\over 2}\right)^2$, hence, assuming $x,d \ll L$, we have $\Delta_{12}=L_1-L_2\approx {xd \over L}$. The constructive interferences are characterized by the waves arriving in phase, therefore by the condition $\Delta_{12}=n\lambda$, where $\lambda$ is the incoming wavelength and $n\in \mathbb{Z}$. Destructive interferences, on the other hand, are characterized by a phase opposition condition, therefore $\Delta_{12}=\left(n+{1\over 2}\right)\lambda$. Hence, the positions of the constructive and destructive interference peaks are given by $x=n{\lambda L\over d}$ and $x=\left(n+{1\over 2}\right){\lambda L\over d}$, respectively. Therefore, if we neglect diffraction phenomena due to the finite size of the slits, all fringes have the same intensity and are separated from each other by the same distance ${\lambda L\over d}$, directly proportional to the wavelength $\lambda$. Note that, in the construction of a wave packet, which involves the superposition of numerous wavelengths, only for the central fringe ($n=0$) will the point of constructive interference be the same for all wavelengths. Therefore, the central fringe will necessarily be the brightest. 

Consider now the situation with three slits; see Figure~\ref{a1}. Reasoning in a similar way as before, and assuming that also the approximation  $d \ll x$ holds (this means that our reasoning will no longer be valid for the central fringe), we have $\Delta_{12}=L_1-L_2\approx {xd \over L}$,  $\Delta_{23}=L_2-L_3\approx {xd \over L}$ and $\Delta_{13}=L_1-L_3=L_1-L_2+L_2-L_3=\Delta_{12}+\Delta_{23}\approx {2xd \over L}$. So, if we call $\psi_1$, $\psi_2$ and $\psi_3$ the three amplitudes at point $x$, associated with the three slits, the intensity $I(x)$ of their superposition is given by the modulus squared of their sum: $I(x)=|\psi_1+\psi_2+\psi_3|^2=I_0|1+e^{ik\Delta_{12}}+e^{ik\Delta_{13}}|^2$, where $k={2\pi\over\lambda}$ is the angular wave number. Using the previous approximations, we can write:  $I(x)\approx I_0|1+e^{ik {xd \over L}}+e^{i2k {xd \over L}}|^2$. Clearly, the peaks of maximum intensity are exactly in the same positions as the two-slit problem. However, the absolute minima are not. Indeed, they correspond to the values $x=\left(n\pm{1\over 3}\right){\lambda L\over d}$, for which one can easily check that $I(x)=0$. In addition to the absolute maximums and minimums, in the three-slit problem there are also local peaks (fringes of lesser intensity); see Figure~\ref{a1}. These appear exactly where, in the two-slit problem, the absolute minima were found. Indeed, for points  $x=n{\lambda L\over d}$, we have $I(x)=I_0$. The intensity of these peaks is therefore one third of that of the maximum peaks. Note that the approximation  $d \ll x$ does not allow to determine the condition for the central fringe, or fringes close to it. However, for $x$ small we can use the approximations $\Delta_{12}\approx {d \over L}(x-{d\over 2})$ and  $\Delta_{23}\approx  {d \over L}(x+{d\over 2})$. Cleary, for symmetry reasons, $\Delta_{13}=0$, for all $\lambda$, but we also have $\Delta_{12} \approx 0$ and $\Delta_{23}\approx 0$, since $d \ll L$, which explains why the central fringe remains the most intense one also in the situation with three slits.

Finally, let us briefly consider the question of the peak width in the single-slit situation. If the slit is of width $a$, then, given a point $x$ on the screen, every point $y\in [-{a\over 2}, {a\over 2}]$ within the slit will contribute with a wave $\psi_y(x)\propto  e^{ik\Delta_y}$, with 
$\Delta_y=L_{-{a\over 2}}-L_y=\sqrt{L^2+\left(x+{a\over 2}\right)^2}-\sqrt{L^2+(x-y)^2}$. Assuming that $a\ll x,L$, we have the approximation $\Delta_y\approx {xy\over L}$ (ignoring terms that do not depend on $y$). So, the intensity $I(x)$, resulting from the superposition of all these contributions is 
\begin{equation}
I(x)=|\int_{-{a\over 2}}^{a\over 2}\psi_y(x) dy|^2\approx I_0|\int_{-{a\over 2}}^{a\over 2} e^{i{kx\over L}y} dy|^2=I_0 \left({\sin{kxa\over 2L}\over{kxa\over 2L}}\right)^2
\label{Fraunhofer}
\end{equation}
This is the classic Fraunhofer diffraction pattern from a rectangular slit, with the central maximum being the most intense and broadest, followed by weaker lateral minima and maxima. Since this function is zero for ${kxa\over 2L}={\pi xa \over\lambda L}=\pm\pi$, the central peak extends from $x=-{\lambda L\over a}$ to $x={\lambda L\over a}$, i.e., its width is $\Delta x={2\lambda L\over a}$.

\end{appendices}


\begin{thebibliography}{99}
\bibitem{thomas1804} Thomas, Y. (1804). The Bakerian lecture. Experiments and calculation relative to physical optics. {\it Philos. Trans. R. Soc. Lond. 94}: pp. 1--16.
 \bibitem{taylor1909} Taylor, G. I. (1909). Interference fringes with feeble light. {\it Proc. Camb. Phil. Soc. 15}, pp. 114--115.
\bibitem{jonsson1961} Jonsson, C. (1961). J\"{o}nsson, C. Elektroneninterferenzen an mehreren k\"{u}nstlich hergestellten Feinspalten. {\it Z. Phys. 161}, pp. 454--474.
\bibitem{merli1974} Merli, P. G.,  Missiroli, G. F. \& Pozzi, G. (1976). ‘On the statistical aspect of electron interference phenomena. {\it Am. J. Phys. 44}, pp. 306--307.
\bibitem{rosa2012} Rosa, R. (2012). The Merli-Missiroli-Pozzi Two-Slit Electron-Interference Experiment. {\it Phys. Perspect. 14}, 178--195.
\bibitem{zeilinger1988} Zeilinger, A., Gähler, G., Shull, C. G. et al. (1988). Single- and double-slit diffraction of neutrons. {\it Rev. Mod. Phys. 6}, pp. 1067--1073.
\bibitem{carnal1991} Carnal, O. \&  Mlynek, J. (1991). Young’s double-slit experiment with atoms: A simple atom interferometer. {\it Phys. Rev. Lett. 66}, pp. 2689--2692.
\bibitem{fein2019}  Fein, Y. Y., Geyer, P., Zwick, P. et al. (2019): Quantum superposition of molecules beyond 25 kDa. {\it Nat. Phys. 15}, pp. 1242--1245.
 \bibitem{jammer1974} Jammer, M. (1974). {\it The Philosophy of Quantum Mechanics: The Interpretations of Quantum Mechanics in Historical Perspective}. New York: John Wiley \& Sons, pp. 109--158.
  \bibitem{feynman1964} Feynman, R. P. , Leighton, R. B. \& Sands, M. (1964), {\it The Feynman Lectures on Physics Volume III}, New York: Addison-Wesley.
\bibitem{aerts2009} Aerts, D. (2009). Quantum particles as conceptual entities: A possible explanatory framework for quantum theory. {\it Found. Sci. 14}, pp. 361--411.

\bibitem{Hampton1988} Hampton, J. A. (1988). Disjunction of natural concepts. {\it Memory \& Cognition 16}, pp. 579--591.
\bibitem{aerts2009b} Aerts, D. (2009). Quantum structure in cognition. {\it J. Math. Psychol. 53}, pp. 314--348.

\bibitem{aertssassoli2018} Aerts, D. and Sassoli de Bianchi, M. (2018). Quantum Perspectives on Evolution. In Shyam Wuppuluri, Francisco Antonio Doria (Eds.), {\it The Map and the Territory: Exploring the Foundations of Science, Thought and Reality}.  Springer: The Frontiers collection, pp. 571--595.
\bibitem{aertssassolisozzoveloz2019} Aerts, D., Sassoli de Bianchi, M., Sozzo, S. and Veloz, T. (2019). From Quantum Axiomatics to Quantum Conceptuality. {\it Act. Nerv. Super. 61}, pp. 76--82.
\bibitem{aertsetal2020} Aerts, D., Sassoli de Bianchi, M., Sozzo, S. and Veloz, T. (2020). On the conceptuality interpretation of quantum and relativity theories. {\it Found. Sci. 25}, 5--54.
\bibitem{sassoli2021} Sassoli de Bianchi, M. (2021). A non-spatial reality. {\it Found. Sci. 26}, pp. 143--170. 
\bibitem{aertssassoli2022}  Aerts, D. and Sassoli de Bianchi, M. (2022). On the irreversible journey of matter, life and human culture. In: Wuppuluri, S., Stewart, I. (Eds.), {\it From Electrons to Elephants and Elections. The Frontiers Collection}. Springer, pp. 821-842.
\bibitem{aertssassoli2024} Aerts, D. and Sassoli de Bianchi, M. (2024). The physics and metaphysics of the conceptuality interpretation of quantum mechanics. \emph{arXiv:2310.10684 [quant-ph]}.
\bibitem{aertssassolisozzoveloz2024a} Aerts, D., Sassoli de Bianchi, M. and Sozzo, S. (2024b). From Quantum Cognition to Conceptuality Interpretation I: Tracing the Brussels Group's Intellectual Journey. To be published in:  \emph{Philos. Trans. R. Soc. A}. arXiv: 2412.06799 [physics.hist-ph].
\bibitem{aertssassolisozzoveloz2024b} Aerts, D., Sassoli de Bianchi, M. and Sozzo, S. (2024c). From Quantum Cognition to Conceptuality Interpretation II: Unraveling the Quantum Mysteries. To be published in: \emph{Philos. Trans. R. Soc. A}. arXiv: 2412.19809 [physics.hist-ph].

\bibitem{Khrennikov2010} Khrennikov, A. Y. (2010). \emph{Ubiquitous Quantum Structure}. Berlin: Springer.
\bibitem{BusemeyerBruza2012} Busemeyer, J. R. \& Bruza, P. D. (2012). \emph{Quantum Models of Cognition and Decision}, Cambridge University Press, Cambridge.
\bibitem{HavenKhrennikov2013} Haven, E. \& Khrennikov, A.Y. (2013). \emph{Quantum Social Science}, Cambridge University Press, Cambridge.
\bibitem{Wendt2015} Wendt, A. (2015). \emph{Quantum mind and social science}, Cambridge University Press; Cambridge.
\bibitem{aertsetal2019} Aerts, D., Aerts Argu\"{e}lles, J., Beltran, L., Geriente, S., Sassoli de Bianchi, M., Sozzo, S. \& Veloz, T. (2019). Quantum entanglement in physical and cognitive systems: a conceptual analysis and a general representation. {\it Eur. Phys. J. Plus. 134}: 493.
\bibitem{aertsbeltran2020} Aerts, D. \& Beltran, L. (2020). Quantum structure in cognition: Human language as a Boson gas of entangled words. {\it Found. Sci. 25}, pp. 755--802.
\bibitem{aertsaerts2022} Aerts, D. \& Aerts Argu\"{e}lles, J. (2022). Human Perception as a Phenomenon of Quantization. {\it Entropy 24}, 1207.

\bibitem{sorkin1994} Sorkin, R. D. (1994). Quantum mechanics as quantum measure theory. {\it Mod. Phys. Lett. A 33}, pp. 3119--3127.
\bibitem{ududec2011} Ududec, C., Barnum, H. \& Emerson, J. (2011). Three Slit Experiments and the Structure of Quantum Theory. {\it Found. Phys. 41}, pp. 396--405.
\bibitem{nyman2011} Nyman, P. \& Basieva, I. (2011). Quantum-Like Representation Algorithm for Trichotomous Observables {\it Int. J. Theor. Phys. 50}, pp. 3864--3881.
\bibitem{sinhaetal2009} Sinha, U., Couteau, C., Medendorp, Z. et al. (2009). Testing Born’s Rule in Quantum Mechanics with a Triple Slit Experiment. {\it AIP Conf. Proc. 1101}, pp. 200--207.
\bibitem{sinhaetal2010} Sinha, U., Couteau, C., Jennewein, T., et al. (2010). Ruling Out Multi-Order Interference in Quantum Mechanics. {\it Science 329}, pp. 418--421.
\bibitem{sinha2011}  Sinha, U. (2011). Born Rule(s). {\it AIP Conference Proceedings 1384}, pp. 254--260.
\bibitem{sollneretal2012} S\"{o}llner, I., Gsch\"{o}sser, B., Mai, P. et al. (2012). Testing Born’s Rule in Quantum Mechanics for Three Mutually Exclusive Event. {\it Found. Phys. 42}, pp. 742--751.
\bibitem{deraedt2012} De Raedt, H. \& Hess, K. (2012). Analysis of multipath interference in three-slit experiments. {\it Phys. Rev. A 85}, 012101.
\bibitem{feynman1965} Feynman R. P. \& Hibbs, A. R. (1965). {\it Quantum Mechanics and Path Integrals}, McGraw-Hill, New York, 1965. 
\bibitem{yabuki1986} Yabuki, H. (1986). Feynman Path Integrals in the Young Double-Slit Experiment. {\it Int. J. Theor. Phys. 25}, pp. 159--174.
\bibitem{sinhatal2015} Sinha, A., Vijay, A. \& Sinha, U. (2015). On the superposition principle in interference experiments. {\it Sci. Rep. 5}, 10304.
\bibitem{quach2017} Quach, J. Q. (2017). Which-way double-slit experiments and Born-rule violation. {\it Phys. Rev. A 95}, 042129.
\bibitem{sinhaetal2014} Sawant, R., Samuel, J., Sinha, A., Sinha, S. \& Sinha, U. (2014). Nonclassical paths in quantum interference experiments. {\it Phys. Rev. Lett. 113}, 120406.
\bibitem{maga2016} Maga\~{n}a-Loaiza, O., De Leon, I., Mirhosseini, M. et al. (2016). Exotic looped trajectories of photons in three-slit interference. {\it Nat. Commun. 7}, 13987.
\bibitem{franson2010} Franson, J. D.  (2010). Pairs Rule Quantum Interference. {\it Science 329}, pp. 396--397.
\bibitem{rengaraj2018} Rengaraj, G., Prathwiraj, U., S. N. Sahoo et al. (2018). Measuring the deviation from the superposition principle in interference experiments. {\it New J. Phys. 20}, 063049. 


\end{thebibliography}
\end{document}